\documentclass[12pt]{article}
\usepackage{amsfonts}
\usepackage{amsmath, amssymb}
\usepackage{youngtab}
\usepackage{hyperref}
\usepackage[dvips]{color}
\usepackage{psfrag}
\usepackage{feynmp}
\usepackage[pdftex]{graphicx}
\usepackage{braket}
\usepackage{bm}
\usepackage{subfigure}

\usepackage{ulem}

\usepackage{comment}
\includecomment{pdffig}
\allowdisplaybreaks[1]

\Yboxdim{5pt}

\makeatletter
    
    \@addtoreset{equation}{section}
  \makeatother

\usepackage{color}

\def\rem#1{}

\renewcommand{\title}[1]{\vbox{\center\LARGE{#1}}\vspace{5mm}}
\renewcommand{\author}[1]{\vbox{\center\large#1}\vspace{5mm}}

\begin{document}
\bibliographystyle{utphys}
\begin{titlepage}
\begin{center}
\vspace{5mm}
\hfill {\tt 
%IPMU21-0003
}\\
\vspace{20mm}

\title{
\LARGE  
't Hooft  surface operators  in five dimensions and  elliptic Ruijsenaars operators 
}
\vspace{7mm}

Yutaka Yoshida

\vspace{6mm}

\vspace{3mm}
Department of Physics, Tokyo Institute of Technology, Tokyo, 152-8551, Japan

\end{center}

\vspace{7mm}
\abstract{
We introduce  codimension three magnetically charged surface operators in five-dimensional (5d) $\mathcal{N}=1$   supersymmetric gauge on $T^2 \times \mathbb{R}^3$. 
We  evaluate the vacuum expectation values (vevs) of surface operators by   supersymmetric localization techniques. Contributions of  Monopole bubbling effects to the path integral    
 are given   by  elliptic genera  of  world volume theories on D-branes.
Our result gives  an elliptic deformation of  the SUSY localization formula \cite{Ito:2011ea} (resp. \cite{Okuda:2019emk, Assel:2019yzd}) of BPS 't Hooft loops (resp. bare monopole operators)  in 4d $\mathcal{N}=2$ (resp. 3d $\mathcal{N}=4$) gauge theories. 
 We define  deformation quantizations of vevs of surface operators in terms of   the Weyl-Wigner transform, where the $\Omega$-background parameter play the role of the Planck constant.
 For 5d   $\mathcal{N}=1^*$ gauge theory, we find that   the deformation quantization  of the surface operators in the anti-symmetric representations agrees with 
the type A elliptic Ruijsenaars operators. The mutual commutativity of  these difference operators  is related to the commutativity of products of 't Hooft surface operators.
}
\vfill

\end{titlepage}
\tableofcontents

\section{Introduction}
An 't Hooft loop  operator \cite{tHooft:1977nqb}  in a four-dimensional (4d) gauge theory is an example of disorder operator defined  by a  boundary condition of  the gauge field with a prescribed   singularity  along the loop. 
In supersymmetric (SUSY) gauge theories,  BPS analogues of  disorder operators preserving a part of the supersymmetry  have interesting properties in a variety of situations. 

In three dimensions,  BPS monopole operators  defined by singular boundary conditions at a point in the spacetime \cite{Borokhov:2002cg, Borokhov:2003yu} play   crucial roles in the study of quantum corrections and dualities in supersymmetric gauge theories. In 3d  $\mathcal{N}=4$ non-abelian gauge theories, the moduli space of Coulomb branch vacua receives non-perturbative corrections from 't Hooft-Polyakov monopoles. 
In general, it is difficult to exactly evaluate the non-perturbative corrections  to the hyperK\"{a}hler metric of the moduli space of the Coulomb branch vacua.  
 An algebra consisting of the vacuum expectation values (vevs) of the monopole operators and Coulomb branch scalars called Coulomb branch chiral 
ring    conjecturally gives the coordinate ring of the moduli space of Coulomb branch vacua  as an algebraic variety.
Here an important point is that the quantum corrections to a Coulomb branch chiral ring   is easier to handle  than   the  corrections to  the hyperK\"{a}hler metric.
In fact,  the Coulomb branch chiral rings in  quiver gauge theories of ADE-type   were  determined  in   \cite{Bullimore:2015lsa}  with  certain assumptions; abelianization and  the mirror symmetry of abelian gauge theories.  In \cite{Okuda:2019emk}, and also \cite{Assel:2019yzd}, the exact computation of vevs of monopole operators was developed in terms of  supersymmetric localization methods. Then  it was found that 
 the algebras of  monopole operators and Coulomb branch obtained by the supersymmetric localization formula agrees with the 
Coulomb branch chiral rings and their deformation quantizations  in \cite{Bullimore:2015lsa, Braverman:2016wma, Braverman:2016aa}

In four dimensions,   BPS 't Hooft loops 
 have attracted a lot of attention from the mathematical physics viewpoints for more than a decade.
For example,  there exists the S-duality between Wilson loops and 't Hooft loops in 4d $\mathcal{N}=4$  super Yang-Mills theory \cite{Kapustin:2005py, Kapustin:2006pk}. In AGT correspondence \cite{Alday:2009aq}, 't Hooft loops in 4d $\mathcal{N}=2$   gauge  theories belonging to the class $\mathcal{S}$ \cite{Gaiotto:2009we}   conjecturally agree with  Verlinde loops in Toda theories \cite{Alday:2009fs, Drukker:2009id,  Gomis:2010kv, Gomis:2011pf,Ito:2011ea}. On the  $\Omega$-background,  the algebra of  the 't Hooft loops, Wilson loops, Dyonic loops and equivalently the algebra of Verlinde loops  gives a deformation quantization of Coulomb branch of 4d $\mathcal{N}=2$ gauge theory on $S^1 \times \mathbb{R}^3$ and also gives a deformation quantization of 
the moduli space of flat connections on a punctured Riemann surface   \cite{Gaiotto:2010be, Nekrasov:2011bc, Dimofte:2011jd, Ito:2011ea}. In this story, 
supersymmetric localization method   provides a powerful method to checks the correspondence between  BPS 't Hooft loops and Verlinde operators.

Although  quantum field  theories in five dimensions  are  non-renormalizable by the power counting argument, some class in 5d $\mathcal{N} =1$  supersymmetric gauge theories have non-trivial fixed points in the renormalization group flow  and make sense as quantum field theories \cite{Seiberg:1996bd, Aharony:1997ju}. Supersymmetric localization formulas of partition functions and supersymmetric indices on five-dimensional manifolds give non-trivial quantitative tests of  predictions for quantum aspects of 5d $\mathcal{N}=1$ supersymmetric theories.
In this paper we introduce  a BPS analogue of disorder operators in 5d $\mathcal{N}=1$ supersymmetric gauge theories on  $T^2 \times \mathbb{R}^3$ by imposing boundary conditions that has  a Dirac monopole singularity extending along a two dimensional torus $T^2$.  The  BPS  disorder operators  which we call as BPS 't Hooft  surface operators  are  five-dimensional analogues of  BPS 't  Hooft loops in 4d $\mathcal{N}=2$ gauge theories  on $S^1 \times \mathbb{R}^3$, and  BPS monopole operators in 3d $\mathcal{N}=4$ gauge theories on $\mathbb{R}^3$. We evaluate the vevs of 't Hooft surface operators by  supersymmetric localization techniques, and study their properties.

This article is organized as follows. In Section \ref{sec:section2}, we introduce the 
BPS 't Hooft surface operators in the path integral formalism by imposing certain boundary conditions for the 
fields in the five-dimensional supermultiplets. In Section \ref{sec:section3}, we study the supersymmetric localization computation of vevs of 't Hooft surface operators and evaluate the classical and  the one-loop contribution to the vevs of 't Hooft surface operators.  In the path integral, there exists non-perturbative corrections coming from the monopole bubbling effect; the path integral over the moduli space of certain monopole solutions. In Section \ref{sec:bubbling}, we evaluate monopole bubbling effect in the vacuum expectation values (vevs) of  't Hooft surface operators in terms of D-brane realizations of monopole bubbling. The monopole bubbling effects contributing to the vevs of 't Hooft surface operators are given by elliptic genera of 
the low energy  world volume theories on D-branes.  In Section \ref{sec:section5}, we define the products of vevs of 
't Hooft surface operators in terms of the Moyal product and also define the deformation quantization 
of vevs of the surface operators in terms of the Weyl-Wigner transform. We find that the deformation quantization of surface operators in 5d $\mathcal{N}=1^*$ gauge theory coincides with 
simultaneously commuting difference operators appearing in an integrable system, called elliptic Ruijsenaars operators. In Section \ref{sec:section6}, we study the algebra of surface operators with respect to 
the Moyal product in 5d $\mathcal{N}=1^*$ gauge theory. In Section \ref{sec:section7}, we discuss our results and future problems.

%%%%%%%%%%%%%%%%%%%%%%%%%%%%%%%%%%%%%%%%%%%%%%%%%%%%%%%%%%%%%%%%%%%%%%%%%%%%%%%%%%
\section{Monopole surface operator in 5d $\mathcal{N}=1$ SUSY gauge theories}
\label{sec:section2}
In this section we explain the decomposition of the ten-dimensional (10d) vector multiplet on the spacetime $\mathbb{R}^{10}$ to 
a 5d vector multiplet and a 5d hypermultiplet in the adjoint representation on the spacetime $\mathbb{R}^{5}$ by the dimensional reduction. Next we
 define the vevs of an 't Hooft surface operator as a  supersymmetric indices on $T^2 \times \mathbb{R}^3$.

%%%%%%%%%%%%%%%%%%%%%%%%%%%%%%%%%%%%%%
\subsection{5d SUSY gauge theory from 10d super Yang-Mills theory}
The convention of the gauge covariant derivative is $D_{M}=\partial_{M}+{\rm i} A_M$. 
The  indices $\mu, \nu, \cdots \in \{ 0, 1, 2, 3, 4 \}$ express the subscripts for the five-dimensional spacetime.
The indices $M, N, \cdots \in \{ 0, 1, \cdots, 9\}$ label the subscripts for the ten-dimensional spacetime.
The spacetimes $\mathbb{R}^5$ and $\mathbb{R}^{10}$  have the Euclidean signature metrics $\delta_{\mu \nu}$ for $\mu,\nu=0,1,2,3,4$ and $\delta_{M N}$ for $M,N=0,1,\cdots,9$, respectively.
$x^{M}$ for $M=0,1,\cdots, 9$ denotes the coordinate of $\mathbb{R}^{10}$. 
The definition and properties of the $16 \times 16$ gamma matrices $\Gamma^{M}$ and $\tilde{\Gamma}^{M}$ are summarized in Appendix \ref{sec:appendix1}.

By the dimensional reduction, the gauge field $A_{M}$ for $M=0,1,\cdots, 9$ and the gaugino $\Psi$ in the 10d maximal super Yang-Mills theory 
 are decomposed to the 5d $\mathcal{N}=1$ supermultiplets as follows.
In the dimensional reduction in the directions $x^{i}$ for $i=5,6,7,8,9$, the five dimensional gauge fields $A_{\mu}$ for $\mu=0, 1, 2, 3, 4$,  an adjoint scalar $\sigma=A_{9}$ and a fermion $\lambda=\frac{1}{2}(1-\Gamma^{5678})\Psi$ form a
5d $\mathcal{N}=1$ vector multiplet.  Scalars $\Phi_{i}:=A_{i}$ for $i=5, 6, 7, 8$  and a fermion $\psi:=\frac{1}{2}(1+\Gamma^{5678})\Psi$ form a 5d $\mathcal{N}=1$ hypermultiplet in the adjoint representation.

The action of the 5d $\mathcal{N}=1$ super Yang-Mills theory is given by 
\begin{align}
S_{\text{vec}}&=\frac{1}{g^2}\int_{ \mathbb{R}^5} d^5 x  \mathrm{Tr} \Bigl[ \frac{1}{2}   F^2_{\mu \nu}+   (D_{\mu} \sigma)^2-\lambda \Gamma^{\mu}  D_{\mu} {\lambda}+{\rm i} \lambda \Gamma^9 [{\lambda},\sigma]   \Bigr].
\label{SYMaction}
\end{align}
Here $g$ is the Yang-Mills coupling constant.  A symbol $\mathrm{Tr}$ is a trace taken over the Lie algebra $\mathfrak{g}$ of the gauge group $G$.  
The action of the 5d hypermultiplet in the adjoint representation is given by
\begin{align}
S_{\text{hyp}}&=\int_{\mathbb{R}^5} d^5 x \mathrm{Tr} \Bigl[   (D_{\mu} \Phi_i)^2-\frac{1}{2} [\Phi_i, \Phi_j]^2 -[\sigma, \Phi_i]^2 
\nonumber \\ &
 \quad \qquad \qquad \qquad 
-\psi \Gamma^{\mu}  D_{\mu} \psi-{\rm i} \psi \Gamma^9 [\sigma, \psi]
-{\rm i} \psi \Gamma^{i} [\Phi_i, \psi]  \Bigr].
\label{hypaction}
\end{align}
We can introduce  a fugacity (mass) $m_{\text{ad}}$ for $U(1)$ flavor symmetry of the adjoint hypermultiplet, which breaks the $\mathcal{N}=2$ supersymmetry to an $\mathcal{N}=1$ supersymmetry  in five dimensions. 
In particular, the 5d $\mathcal{N}=1$ supersymmetry obtained by the mass deformation of 5d $\mathcal{N}=2$ supersymmetry is called 5d $\mathcal{N}=1^*$  supersymmetry.

To apply supersymmetric localization method,  we need at least one off-shell supercharge. For the supersymmetric gauge theories,  by adding  the action of  auxiliary fields $\sum_{i=1}^7K_i^2$ to the above actions \eqref{SYMaction} and \eqref{hypaction},  one can keep some of the  supercharges without using the equation of motion, i.e. off-shell level \cite{Berkovits:1993hx}.  To write a supersymmtry transformation, we choose a supersymmetric variation parameter $\varepsilon$ and introduce  parameters $\nu_j$ for $j=1,\cdots, 7$ given by 
\begin{align}
\varepsilon&=\frac{1}{\sqrt{2}}({1}, 0^{7},1,0^{7})=\frac{1}{\sqrt{2}}({1}, \underbrace{0, \cdots,0}_7,1, \underbrace{0, \cdots, 0}_7) \,, 
\label{eq:SUSYvar}
\\
\nu_{j}
&=\left\{\
\begin{array}{ll}
\displaystyle
\Gamma^{8, j+4} \varepsilon & \text{ for } j=1,2,3 , \\
\displaystyle
 \Gamma^{8 9} \varepsilon & \text{ for } j=4, \\
\Gamma^{8, j-4} \varepsilon & \text{ for } j=5,6,7.
\end{array}
\right. 
\label{eq:tildeGamma}
\end{align}
Then the actions \eqref{SYMaction} and \eqref{hypaction} with the action of auxiliary fields are  invariant under the following  off-shell supersymmetry transformation: 
\begin{align}
\label{eq:SUSYtrans1}
{\sf Q} \cdot A_{M}&= \varepsilon \Gamma_M \Psi\,,  \\
\label{eq:SUSYtrans2}
{\sf Q} \cdot \Psi &= \frac{1}{2} \Gamma^{M N} F_{M N} \varepsilon+{\rm i} K^{i} \nu_i\,, \\
\label{eq:SUSYtrans3}
 {\sf Q} \cdot K_i &= {\rm i} \nu_i \Gamma^{M} D_{M} \Psi\,. 
\end{align}
Here the action of the off shell supercharge ${\sf Q}$  is  defined by $[{\sf Q}, X]$ (resp. $\{ {\sf Q}, X \}$) for the Grassmann even fields $X=A_M, K_i$ (resp. the Grassmann odd field $X=\Psi$).  
The square of the SUSY transformation generates
\begin{align}
{\sf Q}^2 \cdot A_{M}&= -2 F_{\bar{z} M}\,, \\
{\sf Q}^2 \cdot \Psi &= -2 D_{\bar{z} }\Psi\,, \\
 {\sf Q}^2 \cdot K_i &= -2  D_{\bar{z} } K_i\,. 
\end{align}
 Here
 $z=  x^4+{\rm i} x^0$ and $\bar{z}=  x^4 -{\rm i} x^0$.  
If we replace the  representation for the fields in the adjoint hypermultiplet by a symplectic representation $\mathcal{R} \oplus \overline{\mathcal{R}}$ of the gauge group $G$, we obtain the 
supersymmetry transformation and the action for the hypermultiplet in  a symplectic representation $\mathcal{R} \oplus \overline{\mathcal{R}}$.

%%%%%%%%%%%%%%%%%%%%%%%%%%%%%%%%%%%%%%%%%%%%%%
\subsection{Monopole surface operators as SUSY indices on $T^2 \times \mathbb{R}^3$}
We consider a twisted compactification  of 5d supersymmetric gauge theories on the two-dimensional torus $T^2$   defined by 
\begin{align}
T^2&:=\{ (x^4, x^0) | \, x^4+{\rm i} x^0 \equiv  x^4+{\rm i} x^0+2\pi  \equiv x^4+{\rm i} x^0+ 2 \pi  \tau  \} \nonumber \\
&=\{ (z, \bar{z}) | \, z  \equiv z+2\pi  \equiv  z+ 2 \pi  \tau  \}.
\end{align}
Here $\tau=\tau_1+i \tau_2$ is the  moduli of the torus $T^2$.  A symbol  ``$\equiv$'' denotes the  identification. We also introduce another coordinate $0 \le s, t \le 2 \pi$ of 
$T^2$ defined by 
\begin{align}
x^4=s +\tau_1 t, \quad x^0=\tau_2 t.
\end{align} 
We impose  the following twisted boundary condition along the $x^0$ and $x^4$ directions for the fields:
\begin{align}
X(z) \equiv X(z+2\pi), \quad X(z+2 \pi\tau) \equiv e^{-2 \pi {\rm i} \epsilon ({\sf J}_3+{\sf I}_3)} \prod_{f} e^{- 2\pi {\rm i} m_f {\sf F}_f }X(z)
\label{eq:twisedBS}
\end{align} 
where $X(z)$ denotes a field in the 5d supermultiplets.  Here we have suppressed the coordinates dependence on $x^i$ for $i=1,2,3$ and $\bar{z}$    to shorten the notation.

A symbol  ${\sf J}_3$ denotes a generator of the rotation  in the $x^1, x^2$-plane around the origin. When we emphasize  the presence of a fugacity $\epsilon$ called an $\Omega$-background parameter, we 
write a spacetime $\mathbb{R}^3$ in the $x^{1,2,3}$-directions as $\mathbb{R}^2_{\epsilon} \times \mathbb{R}$. 
A symbol ${\sf I}_3$ is a generator of $\mathfrak{u}(1) \subset \mathfrak{su} (2)_H$, where $\mathfrak{su}(2)_H$ is the Lie algebra of  the R-symmetry group  for the 5d $\mathcal{N}=1$ supersymmetry algebra.
 ${\sf F}_f$'s are generators (charges) of the Cartan subalgebra of 
the flavor symmetry group acting on the hypermultiplets in the representation $\mathcal{R} \oplus \overline{\mathcal{R}}$. $m_f$'s are fugacities for these generators. 
 If $\mathcal{R}$ is the adjoint representation, 
we have a single flavor fugacity $m_{\text{ad}}:=m_1$. 

We introduce the vev of a BPS 't Hooft surface operator $S_{{\bm B} }$ for 5d $\mathcal{N}=1$ supersymmetric gauge theory  on $T^2 \times \mathbb{R}^3$ 
as a supersymmetric index:
\begin{align}
\langle S_{\bm B } \rangle&:= \mathrm{Tr}_{\mathcal{H}_{\bm B}} (-1)^{\sf F} e^{-2 \pi {\sf H}} e^{2 \pi {\rm i} \epsilon ({\sf J}_3+{\sf I}_3)} \prod_{f} e^{ 2\pi {\rm i} m_f {\sf F}_f } \, .
\label{eq:operindex}
\end{align}
To be more precise, the vev is defined by  the path integral in the presence of singular monopoles, which will mentioned later.
A magnetic charge ${\bm B}$ is an element of the coweights lattice $\Lambda_{\text{cw}}$ of the Lie algebra $\mathfrak{g}$. 
Since a Weyl group action of ${\bm B}$ defines a same operator, we may assume ${\bm B}$  in $S_{\bm B}$ as a dominant coweight.
$\mathcal{H}_{\bm B}$ is the Hilbert space of the supersymmetric theory on $S^1_{s} \times \mathbb{R}^3$, where $S^1_s$ is the circle in  the $s$-direction.
A symbol ${\sf F}$ denotes the Fermion number operator. 
We take a coordinate $t$ as  the time direction and  define the Hamiltonian ${\sf H}$ by the generator of translation in the direction $t$.  
In the path integral formalism, the vev of an 't Hooft surface operator is given by 
\begin{align}
\langle S_{\bm B} \rangle= \int_{\text{B.C.}}{\mathcal{D}A \mathcal{D} \Psi \mathcal{D} K} \exp \left( -S_{\text{vec}}-S_{\text{hyp}} -S_{\text{b.d}}    \right) \,.
\label{eq:pathfor}
\end{align}
Here we add a   boundary term $S_{\text{b.d}}$  to regularize the singularity coming  from the Dirac monopole, see \eqref{eq:bdryaction}.
In the path integral, the twisted boundary condition \eqref{eq:twisedBS} is given by   the shift of the time derivative:
\begin{align}
\partial_{t} \mapsto  \partial_t -  {\rm i} \epsilon ({\sf J}_3+{\sf I}_3) - {\rm i}\sum_f m_f {\sf F}_f .
\end{align}
Then  the fugacities are given by background gauge field in the path integral formalism.
In the rest of this paper, we include these background gauge fields in the definition of  covariant derivative in the $t$-direction, i.e.,  $D_t=\partial_t +{\rm i} A_{t} -  {\rm i} \epsilon ({\sf J}_3+{\sf I}_3) - {\rm i} m_f {\sf F}_f$.

``{B.C.}''  in \eqref{eq:pathfor} denotes  the following boundary conditions of the fields in the  path integral.
At the infinitesimal neighborhood  of   $(x^0,0,0,0, x^4) $ with $\forall (x^0,x^4) \in T^2$, we impose a boundary condition admitting a singular Dirac monopole with the magnetic charge ${\bm B}$: 
\begin{align}
\sum_{i=1,2,3} A_{i} d x^i \sim \frac{\bm B}{2} (1-\cos \theta) d \phi, \,\,\, \sigma \sim \frac{\bm B}{2 r}, \quad \text{for } r \to 0\,.
\label{eq:BC1}
\end{align}
Here $(r, \theta, \phi)$ is the polar coordinates of the space $(x^1,x^2,x^3)$. 
 In the path integral we also sum over the  boundary conditions, where  the elements of ${\bm B}$  in \eqref{eq:BC1} are  permuted by  an arbitrary Weyl group action of $W_G$.

We  also have to specify the boundary conditions of the fields at a sufficiently large $r$. 
At the spatial infinity, the gauge fields $A_i$ for $i=0,4$ and a scalar $\sigma$  have definite values. We assume these values are in the Cartan subalgebra $\mathfrak{h}$ of $\mathfrak{g}$.
Since  gauge fields $A_i$ for $i=1,2,3$  have a magnetic charge ${\bm B}$ near  the origin $(x^1,x^2,x^3)=(0,0,0)$, 
a natural choice of boundary value $A_{i}$ for $i=1,2,3$ at the spatial infinity is 
\begin{align}
\sum_{i=1,2,3} A_{i} d x^i \sim \frac{\bm B}{2}  (1-\cos \theta) d \phi, \quad \text{for } r \to \infty\,.
\label{eq:BCinf}
\end{align}
But an important  point here is that we have to take into account not only the boundary condition \eqref{eq:BCinf}, but all the boundary conditions associated with monopole bubbling   (a.k.a. monopole screening )  \cite{Kapustin:2006pk}.
A monopole bubbling  is a phenomenon that an 't Hooft-Polyakov monopole screen  out the charge of the Dirac monopole ${\bm B}$ and  reduced it  to ${\bm p}$  with $||{\bm p}|| < ||{\bm B} ||$, 
 when a smooth 't Hooft-Polyakov monopole  with a magnetic charge ${\bm p} -{\bm B}$ in the coroot lattice $\Lambda_{\text{cr}}$ exists  at the infinitesimal neighborhood of the center of the singular Dirac monopole
with the charge ${\bm B}\in \Lambda_{\text{cw}}$. Here $||{\bm B}||=\sqrt{\mathrm{Tr}({\bm B}^2)}$. 

Monopole bubbling effects in the SUSY localization formula of  BPS 't Hooft loops for $U(N)$ and $SU(N)$ gauge theories were originally studied in \cite{Gomis:2011pf, Ito:2011ea} and were further studied  in \cite{Brennan:2018rcn}.
It turned out that 't Hooft loops agree with Verlinde loop operators by including monopole bubbling effects.  Moreover monopole bubbling effects  are also necessary  to reproduce the correct properties of operator product expansions (OPEs) of BPS 't Hooft loops \cite{Kapustin:2006pk, Ito:2011ea, Hayashi:2019rpw, Hayashi:2020ofu}. In three dimensions, monopole bubbling effects for BPS monopole operators on $\mathbb{R}^3$ were studied in \cite{Okuda:2019emk, Assel:2019yzd}.  It found that  monopole bubbling effects are necessary in order for monopole operators with higher charges  to be generated by the products of  operators with smaller charges. 
From these observations, we  consider all the boundary conditions with monopole bubblings for the surface operator specified by
\begin{align}
\sum_{i=1,2,3} A_{i} d x^i \sim \frac{\bm p}{2} (1-\cos \theta) d \phi, \quad \text{for } r =\text{finite}\,,
\label{eq:BCinf2}
\end{align}
if  ${\bm p}  \in {\bm B} +\Lambda_{\text{cr}}$ with $||{\bm p}|| < ||{\bm B} ||$ exists.
We will study the contribution 
from   the moduli of monopole bubbling  to the path integral in terms of brane constructions in Section \ref{sec:bubbling}.

%%%%%%%%%%%%%%%%%%%%%%%%%%%%%%%%%%%%%%%%%%%%%%%%%%%%
\section{SUSY localization of surface operator}
\label{sec:section3}
In this section,
we perform the path integral for the vev of surface operators in terms of supersymmetric localization. 
Before we explain  the technical details of the localization computation,
 we first summarize  the result of our supersymmetric localization formula. 
We consider that the gauge group is $G$  and   $N_F$ hypermultiplets are in a symplectic  representation $\mathcal{R} \oplus \overline{\mathcal{R}}$ of the gauge group.
The localization formula for the vev of monopole surface operator  is given by
\rem{
\begin{align}
\langle S_{{\bm B}  } \rangle&= \sum_{ {\bm p} \in  W_{G} \cdot {\bm B} }  e^{ {\bm p} \cdot {\bm b}    }
Z^{\text{5d}}_{1\text{-loop}} ( {\bm a} , {\bm m}, {\bm p} ,\epsilon, \tau)  \nonumber \\
&\qquad + \sum_{ {\bm p} \in {\bm B}+\Lambda_{\text{cr}}(\mathfrak{g}) \atop ||{\bm p}|| < ||{\bm B} ||}  e^{ {\bm p} \cdot {\bm b}    }
Z^{\text{5d}}_{1\text{-loop}} ( {\bm a} , {\bm m}, {\bm p} ,\epsilon, \tau) Z_{\text{mono}} ( {\bm a} ,{\bm m}, {\bm p}, {\bm B}, \epsilon, \tau).
\label{eq:localizationfm}
\end{align}
}
\begin{align}
\langle S_{{\bm B}  } \rangle&=  \sum_{ {\bm p} \in {\bm B}+\Lambda_{\text{cr}} \atop ||{\bm p}|| \le ||{\bm B} ||}  e^{ {\bm p} \cdot {\bm b}    }
Z^{\text{5d}}_{1\text{-loop}} ( {\bm a} , {\bm m}, {\bm p} ,\epsilon, \tau) Z_{\text{mono}} ( {\bm a} ,{\bm m}, {\bm p}, {\bm B}, \epsilon, \tau).
\label{eq:localizationfm}
\end{align}
 $Z_{\text{mono}}$'s in \eqref{eq:localizationfm} are the contributions from the monopole bubbling effects. Note that $Z_{\text{mono}}(  {\bm p}, {\bm B})=1$ for $||{\bm p}|| = ||{\bm B} ||$. If ${\bm p} \in {\bm B}+\Lambda_{\text{cr}}$ with $ ||{\bm p}|| < ||{\bm B} ||$ does not 
exist, the localization formula is completely determined  by the  one-loop computations for $||{\bm p}|| = ||{\bm B} ||$.
The right hand side of \eqref{eq:localizationfm} is given as follows.
${\bm b}$ is  defined in \eqref{eq:defb}. 
A pairing ${\bm p} \cdot {\bm b}$ of ${\bm p} $ and $ {\bm b}$  is induced by the trace over $\mathfrak{g}$:
\begin{align}
{\bm p} \cdot {\bm b}=\mathrm{Tr}({\bm p }{\bm b} )\,.
\end{align}
$Z^{\text{5d}}_{1\text{-loop}} ( {\bm a} , {\bm m}, {\bm p}, \epsilon, \tau)$ is the one-loop determinant of the Q-exact action around the saddle point specified by a magnetic charge ${\bm p}$, and  factorizes to
 the one-loop determinant of the 5d $\mathcal{N}=1$ vector multiplet  $Z^{\text{5d.vec}}_{1\text{-loop}}$ and the one of the 5d $\mathcal{N}=1$ hypermultiplet  $Z^{\text{5d.hyp}}_{1\text{-loop}}$:
\begin{align}
Z^{\text{5d}}_{1\text{-loop}} ( {\bm a},  {\bm m}, {\bm p}, \epsilon, \tau) 
=Z^{\text{5d.vec}}_{1\text{-loop}} ( {\bm a}, {\bm p}, \epsilon,  \tau)Z^{\text{5d.hyp}}_{1\text{-loop}}
({\bm a}, {\bm m}, {\bm p}, \epsilon, \tau),
\label{eq:lopptot}
\end{align}
where
\begin{align}
&Z^{\text{5d.vec}}_{1\mathchar `-\text{loop}}({\bm a},\bm{p}, \epsilon, \tau)= 
 \left[\prod_{\alpha \in \mathrm{rt}(\mathfrak{g}) } \prod_{k=0}^{|\alpha \cdot \bm{p}|-1}  
\frac{ \vartheta_1 \left ( \alpha \cdot {\bm a} + \left( \frac{|\alpha \cdot \bm{p}| }{2}-k \right) \epsilon; \tau \right)}{\eta(\tau)} \right] ^{-\frac{1}{2}},
\label{eq:1loopvec}
\\
&Z^{\text{5d.hyp}}_{1 \mathchar `-\text{loop}}({\bm a}, {\bm m}, {\bm p}, \epsilon, \tau)=  
   \left[\prod_{{w} \in \Delta (\mathcal{R})} \prod_{f=1}^{N_F} \prod_{k=0}^{|{ w} \cdot \bm{p}|-1} \frac{ \vartheta_1 \left ( {w} \cdot {\bm a} - { m_f} + \left( \frac{|{ w} \cdot \bm{p}| -1}{2}-k \right) \epsilon; \tau \right)}{\eta(\tau)} \right]^{\frac{1}{2}}, 
\label{eq:1loophy}
\end{align}
$\mathrm{rt}(\mathfrak{g})$ is the set of roots  of $\mathfrak{g}$ and $\Delta(\mathcal{R})$ is  the set of  weights of a representation $\mathcal{R}$.

A theta function $\vartheta_1 \left ( u; \tau \right)$ and eta function $\eta (\tau)$ are defined by
\begin{align}
\vartheta_1 ( u)= \vartheta_1 \left ( u; \tau \right)& := 2 e^{ \frac{ {\rm i} \pi}{4}  \tau} \sin (\pi u) \prod_{i=1}^{\infty} (1-e^{2 \pi {\rm i} n \tau}) (1-e^{2 \pi {\rm i} n \tau} e^{2 \pi {\rm i}  u}) (1-e^{2 \pi {\rm i} n \tau} e^{-2 \pi {\rm i}  u}) \,,
\\
\eta (\tau)&:=e^{\frac{\pi {\rm i} \tau} {12}} \prod_{n=1}^{\infty} (1-e^{2 \pi {\rm i} n \tau})\,.
\end{align}
To shorten expressions, we introduce the following  notation for the theta function: 
\begin{align}
\vartheta_1(\pm x+y)&=\prod_{\alpha=\pm1}\vartheta_1(\alpha x+y),  \quad
\vartheta_1(\pm x \pm y+w)=\prod_{\alpha, \beta=\pm1}\vartheta_1(\alpha x+ \beta y +w).
\end{align}

$Z_{\text{mono}} ( {\bm p}, {\bm B})$ in \eqref{eq:localizationfm}   is interpreted as a contribution from  the path integral over the moduli space of  Bogomolny  equation \eqref{eq:Bog} with  a monopole bubbling  explained in Section \ref{sec:bubbling}.  We gives  explicit computations of $\langle S_{\bm B} \rangle$ for small magnetic charges in Section \ref{sec:section6}.

%%%%%%%%%%%%%%%%%%%%%%%%%%%%%%%%%%%%%%%%%%%%%%%%%
\subsection{Zero locus of Q-exact term in 5d $\mathcal{N}=1$ SUSY gauge theory}
To apply supersymmetric localization procedure, we introduce  the Faddeev–Popov ghosts $c,\bar{c},$  the Nakanishi-Lautrup B-field $\bar{b}$, and  a BRST charge ${\sf Q}_B$.
We will explain  the definition of the BRST charge ${\sf Q}_B$ and the gauge fixing term  in the next subsection. 
We add  one-prameter family of Q-exact action  ${\rm t} \widehat{\sf Q} \cdot V$ to the original action,  and take a   limit ${ {\rm t} \to \infty}$:
\begin{align}
\langle S_{\bm B} \rangle=\lim_{{\rm t} \to  \infty} \int \mathcal{D}A \mathcal{D} \Psi \mathcal{D} K \mathcal{D} c \mathcal{D} \bar{c} \mathcal{D} \bar{b} \exp 
\left( {-S_{\text{vec}}-S_{\text{hyp}} -S_{\text{b.d}}-{\rm t} \widehat{\sf Q} \cdot V }  \right)\,,
\label{eq:path2}
\end{align}
with 
\begin{align}
\widehat{\sf Q}:={\sf Q}+{\sf Q}_B\,.
\end{align}
Then the principle of supersymmetric localization  \cite{Pestun:2007rz} tells us that   the path integral in \eqref{eq:pathfor} is exactly evaluated  in terms of  the one-loop integral around $\widehat{\sf Q} \cdot V=0$ and integrals over  certain moduli spaces of  equations  associated with $\widehat{\sf Q} \cdot V=0$ in \eqref{eq:path2}. 

In this section we focus on the following matter part in  the Q-exact action  $\widehat{\sf Q} \cdot V 
$ : 
\begin{align}
{\sf Q}  \cdot (\Psi, \overline{{\sf Q} \cdot \Psi}) \,, \label{eq:Qexact1}
\end{align}
and study  the zero locus (saddle point locus) of \eqref{eq:Qexact1}.
Here $\overline{{\sf Q} \cdot \Psi}$ is the  complex conjugate of  ${\sf Q}  \cdot \Psi$.   $\left( A , B \right)=\int_{T^2 \times \mathbb{R}^3} A B$, where   the   spinor indices  and Lie algebra indices  in five dimensional fields $A$ and $B$ are contracted.
To write down explicitly the saddle equation ${\sf Q} \cdot \Psi=0$  in \eqref{eq:Qexact1},  
it is convenient to decompose the  fermion $\Psi$ with the sixteen components into Grassmann odd  functions $\Psi_M$ for $M=1,\cdots,9$ and $\Upsilon^{i}$ for $i=1,\cdots,7$ 
as
\begin{align}
 \Psi &=  \sum_{M=1}^{9} \Psi_M \tilde{\Gamma}^{M} {\varepsilon}+{\rm i} \sum_{i=1}^7 \Upsilon^{i} \nu_i\,.
\end{align}
 Here $\Psi_M$ and $\Upsilon^{i}$ are defined by
\begin{align}
  \Psi_M &= \varepsilon \Gamma^M \Psi, \quad
{\rm i} \Upsilon^{i} =\bar{\nu}_i \Psi \,,
\end{align}
with  $\bar{\nu}_i=- \nu_i$. 
Then, we have the supersymmetry transformation of $\Psi_M$ and $\Upsilon_i$ as follows.
\begin{align}
{\sf Q}  \cdot \Psi_M &=\varepsilon \Gamma^M {\sf Q}  \cdot \Psi 
={\rm i} F_{M 0}  +F_{M 4}\,.
\label{eq:QPsi}
\end{align}
Here we used the following properties; $\Gamma^M  \Gamma^{ N L }=\Gamma^{[M}  \Gamma^{ N L] }+2 \delta^{M [N} \Gamma^{ L ]}$, $\varepsilon \Gamma^{[M}  \Gamma^{ N L] } \varepsilon=0$
and $(\varepsilon \Gamma^{ 0 } \varepsilon, \varepsilon \Gamma^{ 1 } \varepsilon, \cdots, \varepsilon \Gamma^{ 9 } \varepsilon) =({\rm i}, 0^3,1,0^5)$.
\begin{align}
{\rm i} {\sf Q} \cdot \Upsilon_{i} &=\bar{\nu}_i {\sf Q} \cdot \Psi \nonumber \\
&=\frac{1}{2} \left( \sum_{j ,k=1}^3 F_{j k} \bar{\nu}_i  \Gamma^{ j k } \varepsilon  + 2 \sum_{j=1}^3 \sum_{ k=5 }^8  F_{j k} \bar{\nu}_i  \Gamma^{ j k } \varepsilon 
+ 2 \sum_{ j=1 }^3  F_{j 9} \bar{\nu}_i  \Gamma^{ j 9 } \varepsilon+ 2 \sum_{ j=5 }^8  F_{j 9} \bar{\nu}_i  \Gamma^{ j 9} \varepsilon \right) \nonumber \\
&\qquad + {\rm i} K^{i} \,. 
\end{align}
Here we used $\bar{\nu}_i \nu_j=-\delta_{ i j}$ and  $\bar{\nu}_i \Gamma^{M 4} \varepsilon=0$ for $i =1 \cdots, 7$ and $M=0,1,\cdots, 9$. Then
\begin{align}
{\rm i} {\sf Q} \cdot \Upsilon_{i =1,2,3} 
&= - \frac{1}{2}  \sum_{j ,k=1}^3 \epsilon_{i jk}    F_{j k} 
+   D_{i} \sigma   - {\rm i} K^{i}. 
\label{eq:saddlemono}
\\
{\rm i} {\sf Q} \cdot \Upsilon_{i=4,5,6,7} 
&=\sum_{j=1}^3 \sum_{ k=5 }^8  D_{j }  \Phi_k \bar{\nu}_i  \Gamma^{ j k } \varepsilon 
+  \sum_{ j=5 }^8  {\rm i} [ \Phi_{j}, \sigma] \bar{\nu}_i  \Gamma^{ j 9} \varepsilon - {\rm i} K^{i}
\label{eq:saddlemat}
\end{align}
Here we used 
\begin{align}
&\bar{\nu}_{i} \Gamma^{j k} \varepsilon=0 \quad \text{for} \quad i \in  \{5, 6, 7 \}, \,\,j, k \in \{ 1, 2, 3\} \\
 &\bar{\nu}_i  \Gamma^{ j 9 } \varepsilon=0  \quad \text{for} \quad i \in \{ 4,5,6,7 \}, \, \, j \in \{ 1,2,3 \}.
\end{align}
From \eqref{eq:QPsi} and \eqref{eq:saddlemono}, we find that the saddle point equation ${\sf Q}  \cdot \Psi=0$ for the bosonic fields in the 5d vector multiplet are decomposed to
\begin{align}
\label{eq:flat}
F_{0 4}&=0 \,, \\
\label{eq:Bog}
D_i \sigma&=\frac{1}{2} \epsilon_{i j k}F_{ j k} \, \, \, (i,j,k=1,2,3)  \,, \\
K_i&=0 \,.
\end{align}
From \eqref{eq:saddlemat}, the  saddle point equation ${\sf Q}  \cdot \Psi=0$ for the 5d hypermultiplet \eqref{eq:saddlemat} is written as 
\begin{align}
\sum_{i=1}^3 \sigma^i D_i q+ [\sigma, q]=0  \,,
\end{align}
where $\sigma^i, i=1,2,3$ are the Pauli matrices, and $q=(q_1,q_2)^T$ is defined by
\begin{align}
q_1:= \Phi_5-{\rm i} \Phi_6+{\rm i} \Phi_7+\Phi_8, \quad q_2:={\rm i} \Phi_5- \Phi_6-\Phi_7-{\rm i} \Phi_8 \,.
\end{align}
The invariance under the flavor symmetry rotation require the saddle point value of $\Phi_i$ for $i=5,\cdots,9$ is zero.

We evaluate the saddle point value of the super Yang-Mills action and the boundary term.
$F_{04}=0$ means that  the saddle point configuration for the gauge fields $A_i$ for $i=0,4$  are  flat connections which are fixed by the value at $r \to \infty$. Let  $\bar{A}_i$, $i=0,4$  be the boundary values $A_{i} |_{r = \infty}=\text{constant}$ for $i=0,4$ and  define ${\bm a}$ as a holomorphic combination of the constant gauge fields: 
\begin{align}
{\bm a}:=\bar{A}_0+{\rm i} \bar{A}_4 \in \mathfrak{h} \otimes {\mathbb{C}}\,.
\label{eq:Flatcon}
\end{align}

Next let $(\bar{A}_{i}, \bar{\sigma})$ for $i=1,2,3$ be a solution  of  the Bogomolny equation \eqref{eq:Bog}.  From the boundary condition of the path integral \eqref{eq:BC1},
$(\bar{A}_{i}, \bar{\sigma})$ behaves as
\begin{align}
\sum_{i=1,2,3} \bar{A}_{i} d x^i &\sim \frac{\bm B}{2} (1-\cos \theta) d \phi, \quad
\bar{\sigma} \sim\frac{{\bm B}}{2r}, \quad \text{for } r \to 0 \,,
\label{eq:BCr0} \\
\sum_{i=1,2,3} A_{i} d x^i &\sim \frac{\bm p}{2} (1-\cos \theta) d \phi, \quad 
\bar{\sigma} \sim\frac{{\bm p}}{2r}+\sigma_0 \quad \text{for } r =\text{finite}\,.
\label{eq:BCrinf}
\end{align}

Here we assume that $\sigma_0$ is a Cartan  valued constant. 
We substitute a saddle point value $(\bar{A}_{i=0,\cdots,4}, \bar{\sigma})$ into the 5d super Yang-Mills action. Then saddle point value of the 5d super Yang-Mills action is given 
\begin{align}
S_{\text{vec}}|_{\text{saddle}}&=\frac{1}{g^2}\int_{T^2 \times ( \mathbb{R}^3 \backslash B^3_{R}) }  d^5 x \mathrm{Tr} \Bigl[ \frac{1}{2}  F^{\mu \nu} F_{\mu \nu}+  D^{\mu} \sigma D_{\mu} \sigma-\lambda \Gamma^{\mu}  D_{\mu} {\lambda}+{\rm i} \lambda \Gamma^9 [{\lambda},\Phi_9]   \Bigr] \Big|_{\text{saddle}} \nonumber \\
&=
\frac{2 \pi \text{vol}(T^2)}{g^2 R } \mathrm{Tr}( {\bm p}^2 )\,.
\label{eq:saddleYM}
\end{align}
where 
\begin{align}
B^3_{R}:=\{ (x^1,x^2,x^3) | \,\, |x^1|^2+ |x^2|^2+ |x^3|^2 \le R^2 \}\,.
\end{align}
In \eqref{eq:saddleYM}, we removed the three-dimensional  ball $B^3_{R}$ from  the  integration region   to regularize the divergence coming from the center of the Dirac monopole.
The divergence  coming from $R \to 0$ is compensated by the saddle point value of the boundary  term $S_{\text{b.d}}$ \cite{Ito:2011ea} as follows. 
\begin{align}
S_{\text{b.d}}|_{\text{saddle}}&=\frac{2}{ g^2} \int_{T^2 \times \partial B^3_{R}}  \text{dvol}(T^2)  \mathrm{Tr} ({\sigma} F_A) \Big|_{\text{saddle}} \nonumber \\
&=-\frac{1}{ g^2} \text{vol}(T^2) 
\mathrm{Tr} \left( 2 \pi \frac{\bm p^2}{R} +4 \pi^2 R {\sigma}_0 {\bm p} \right) \,. 
\label{eq:bdryaction}   
\end{align}
Here $\text{dvol}(T^2)$ is the volume form of $T^2$, and $F_A$ is the restriction of the field strength $\frac{1}{2}\sum_{i, j=1}^3 F_{i j} dx^i \wedge d x^j$ on the boundary $\partial B^3_R$.
 The the saddle point value of the sum of the super Yang-Mills action and the boundary term has a finite value:
\begin{align}
(S_{\text{vec}}+S_{\text{b.d}})|_{\text{saddle}}&=-\frac{4 \pi^2 R}{ g^2}\text{vol}(T^2) 
\mathrm{Tr} \left( {\bm p}  {\sigma}_0 \right) =: {{\bm p}\cdot {\bm b}}\,,    
\end{align}
with
\begin{align}
{\bm b}=-\frac{4 \pi^2 R}{ g^2}\text{vol}(T^2)  {\sigma}_0. 
\label{eq:defb}
\end{align}
On the other hand, the saddle point values of the fields in the hypermultiplet are zero, which means that  the saddle point values of  the action of  hypermultiplet is zero.
 
%%%%%%%%%%%%%%%%%%%%%%%%%%%%%%%%%%%%%%%%%%%%%%%%%%%%%%%%%%
\subsection{Gauge fixing term and BRST transformation}
To evaluate the one-loop determinant, we introduce the action for the ghosts and the gauge fixing term: 
\begin{align}
S_{\text{g.f}}&=\int_{T^2 \times \mathbb{R}^3} d^5 x {\sf Q}_B \cdot \mathrm{Tr} \left( \bar{c} \left(\sum_{i=1,2,3,9} \bar{D}_{M} \tilde{A}_M +\frac{\xi}{2} \bar{b} \right) \right) \nonumber \\
&= \int_{T^2 \times \mathbb{R}^3} d^5 x  \mathrm{Tr} \left(-{\rm i} \bar{c} \bar{D}_{M} D_{M} c+ \bar{b} \left( {\rm i} \bar{D}^{M} \tilde{A}_M+\frac{\xi}{2} \bar{b} \right) \right)
\end{align}
Here the $\mathfrak{g}$-valued ghost fields $c, \bar{c}$ are Grassmann odd, and the $\mathfrak{g}$-valued B-field $\bar{b}$ is a Grassmann even.
$\bar{D}_{M}=\partial_{M}+{\rm i}\bar{A}_M$, where $\bar{A}_M$ denotes the saddle point values of the gauge fields and the vector multiplet scalar given by \eqref{eq:Flatcon}, \eqref{eq:BCr0}, \eqref{eq:BCrinf}, and 
$\bar{A}_{i=5,6,7,8}=0$. $\tilde{A}_M$ is the fluctuation of around $\bar{A}_M$ mentioned in the beginning of Section \ref{eq:1loopdet}.
We define the BRST transformation  by 
\begin{align}
{\sf Q}_B \cdot A_M&=D_{M}c, \quad {\sf Q}_B \cdot \Psi= -{\rm i} [c, \Psi], \quad 
{\sf Q}_B \cdot K_i=-{\rm i} [c, K_i], \nonumber \\
{\sf Q}_B \cdot c&=-\frac{\rm i}{2} [c,c], \quad {\sf Q}_B \cdot \bar{c}=\bar{b}, \quad {\sf Q}_B \cdot \bar{b}=0.
\end{align}
Note that the BRST charge is nilpotent $\{{\sf Q}_B, {\sf Q}_B\}=0$.
The supersymmetry transformation for $(c,\bar{c},\bar{b})$ is defined by 
\begin{align}
{\sf Q} \cdot c&=-{\rm i} \tilde{A}_0-\tilde{A}_4\,, \nonumber \\
  {\sf Q} \cdot \bar{c}&=0\,, \nonumber \\
 {\sf Q} \cdot \bar{b}&=-({\rm i} D_0+D_4) \bar{c} \,.
\end{align}
We define the Q-exact term  by
\begin{align}
\widehat{\sf Q} \cdot V:=\widehat{\sf Q}\left[
 (\Psi, \overline{ \widehat{\sf Q} \cdot \Psi}) +V_{\text{g.f}} \right]
\end{align}
with $\widehat{\sf Q}={\sf Q}+{\sf Q}_B$.

%%%%%%%%%%%%%%%%%%%%%%%%%%%%%%%%%%%%%%%%%%%%%%%%%%%%%%%%%%%%%%%%%%
\subsection{Evaluation of one-loop determinants}
\label{eq:1loopdet}
In the localization computation of the path integral in \eqref{eq:path2}, We decompose the fields 
to $X=\bar{X}+\tilde{X}/ \sqrt{\sf t}$, where $\bar{X}$ denotes the fields satisfy $\widehat{\sf Q} \cdot V=0$, and $\tilde{X}$ is called as the fluctuation fields around the saddle point value.
In the limit ${\sf t} \to \infty$, the quadratic part of the  fluctuation fields in the Q-exact action only 
contributes to the path integral, and the higher order interaction terms in the Q-exact action are negligible, i.e., the one-loop computation for  the fluctuation fields gives the exact answer. Then we may take 
the first order approximation of $\widehat{\sf Q} \cdot \tilde{X}$ in the Q-exact action with respect to  the fluctuation fields.

In order to evaluate the one-loop determinant  in terms of the  index theorem of the transversally elliptic operators,  let us define collections of the fluctuation fields ${\bm X}_i$, $i=0,1$ by 
\begin{align}
{\bm X}_0&=(X_{0,1}, X_{0,2}, \cdots, X_{0,9})=(\tilde{A}_1, \tilde{A}_2,\cdots, \tilde{A}_9)\,, \nonumber\\
 {\bm X}_1&=(X_{1,1}, X_{1,2}, \cdots, X_{1,9})=(\Upsilon_1,\cdots,\Upsilon_7, c, \bar{c})\,.
\end{align}
To make the expression concise, we omitted $\tilde{}$ for the fluctuation fields $\Upsilon_i$ and the ghosts. Note that the saddle point configuration for the ghost and B-field is trivial.
The square of $\widehat{\sf Q}$-transformation generates
\begin{align}
\widehat{\sf Q}^2&=-\frac{\rm i}{\tau_2}\left( \partial_{t}- \tau \partial_s + {\rm i} {\bm a} -{\rm i} \epsilon ({\sf J}_3+{\sf I}_3) - {\rm i} \sum_f m_f {\sf F}_{f} \right)\,.
\label{eq:squareQ} 
\end{align}
At the quadratic order of ${\bm X}_i$, $V$ can be written as
\begin{align}
V&=(\Psi, \overline{ \widehat{\sf Q} \cdot \Psi}) +V_{\text{g.f}} \nonumber \\
&=\left( \left(
    \begin{array}{cc}
            \widehat{\sf Q} \cdot {\bm X}_0, & {\bm X}_1
    \end{array}
  \right),  
 \left(
    \begin{array}{cc}
       D_{0 0} & D_{0 1} \\
      D_{1 0} & D_{1 1}
    \end{array}
  \right) 
\left(
    \begin{array}{c}
        {\bm X}_0   \\
        \widehat{\sf Q} \cdot {\bm X}_1
    \end{array}
  \right)  \right)\,,
\label{eq:Qexacttot}
\end{align}
where $D_{i j}$ for $i,j=0,1$ are linear differential operators which depend on the saddle point field configuration, but are independent of the fluctuations field ${\bm X}_i$. 
Then the Q-exact action $\widehat{\sf Q} \cdot V$ at the quadratic order is given by
\begin{align}
\widehat{\sf Q} \cdot V
&=\left( \left(
    \begin{array}{cc}
       {\bm X}_0 &  \widehat{\sf Q} \cdot {\bm X}_1
    \end{array}
  \right) , \left(
    \begin{array}{cc}
       -\widehat{\sf Q}^2 & 0 \\
      0 & 1
    \end{array}
  \right)  
 \left(
    \begin{array}{cc}
       D_{0 0} & D_{0 1} \\
      D_{1 0} & D_{1 1}
    \end{array}
  \right) 
\left(
    \begin{array}{c}
        {\bm X}_0   \\
        \widehat{\sf Q} \cdot {\bm X}_1
    \end{array}
  \right)  \right) \nonumber \\
&+\left( \left(
    \begin{array}{cc}
      \widehat{\sf Q} \cdot {\bm X}_0 & {\bm X}_1
    \end{array}
  \right) ,  
 \left(
    \begin{array}{cc}
       D_{0 0} & D_{0 1} \\
      D_{1 0} & D_{1 1}
    \end{array}
  \right)
\left(
    \begin{array}{cc}
       -1 & 0 \\
      0 & - \widehat{\sf Q}^2
    \end{array}
  \right) 
\left(
    \begin{array}{c}
        \widehat{\sf Q} \cdot {\bm X}_0   \\
        {\bm X}_1
    \end{array}
  \right)  \right)\,.
\end{align}
Following the argument in \cite{Pestun:2007rz}, one can show that the one-loop determinant is reduced to 
the ratio of the functional determinants over the spaces $\mathrm{Det}_{\mathrm{coker}(D_{10}) }$ and 
$\mathrm{Det}_{\mathrm{ker}(D_{10}) }$:
\begin{align}
Z^{\text{5d}}_{1\text{-loop}}
&=
\left[
\frac{ 
\mathrm{Det} \left(
 \left(
    \begin{array}{cc}
       D_{0 0} & D_{0 1} \\
      D_{1 0} & D_{1 1}
    \end{array}
  \right)
\left(
    \begin{array}{cc}
      - 1 & 0 \\
      0 & -\widehat{\sf Q}^2
    \end{array}
  \right) \right) }
{ \mathrm{Det} \left( 
\left(
    \begin{array}{cc}
      - \widehat{\sf Q}^2& 0 \\
      0 & 1
    \end{array}
  \right)  
 \left(
    \begin{array}{cc}
       D_{0 0} & D_{0 1} \\
      D_{1 0} & D_{1 1}
    \end{array}
  \right) \right) }  
\right]^{\frac{1}{2}} 
= 
\left[
\frac{\mathrm{Det}_{\mathrm{coker}(D_{10}) } \widehat{\sf Q}^2 }{\mathrm{Det}_{\mathrm{ker}(D_{10}) } \widehat{\sf Q}^2 }
\right]^{\frac{1}{2}}\,.
\end{align}
Here $\mathrm{ker}(D_{10})$ is the space of the kernel of the differential operator $D_{10}$, and
$\mathrm{coker}(D_{10})$ is the space of the cokernel of the differential operator $D_{10}$. 
Since
 the differential operator $D_{1 0}$ is decomposed to $D_{10}=D_{\text{vec}} \oplus D_{\text{hyp}}$, where $D_{\text{vec}}$ acting on the fields in the vector multiplet and  $D_{\text{hyp}}$ acting on the ones in the 
hypermultiplet,  the one-loop determinant factorizes to  the product of the one-loop determinants  of the vector multiplet and the hypermultiplet:
\begin{align} 
Z^{\text{5d}}_{1\text{-loop}}=Z^{\text{5d.vec}}_{1\text{-loop}}
Z^{\text{5d.hyp}}_{1\text{-loop} }\,,
\end{align}
with
\begin{align} 
Z^{\text{5d.vec}}_{1\text{-loop}}& =
\left[
\frac{\mathrm{Det}_{\mathrm{coker}(D_{\text{vec}}) } \widehat{\sf Q}^2}{\mathrm{Det}_{\mathrm{ker}(D_{\text{vec}}) } \widehat{\sf Q}^2 }
\right]^{\frac{1}{2}}, \quad
Z^{\text{5d.hyp}}_{1\text{-loop} }=
\left[
\frac{\mathrm{Det}_{\mathrm{coker}(D_{\text{hyp}}) } \widehat{\sf Q}^2} {\mathrm{Det}_{\mathrm{ker}(D_{\text{hyp}}) } \widehat{\sf Q}^2}
\right]^{\frac{1}{2}}\,.
\end{align}
The important point here is that  the ratio of the functional determinants is evaluated in term of   the  index theorem of  transversally elliptic operators. 
Let $\mathrm{ind}(D)$ be the equivariant index of a  transversally elliptic operator  $D= D_{\text{vec}} \text{ or } D_{\text{hyp}}$ defined by
\begin{align} 
\mathrm{ind}(D) =\mathrm{tr}_{\mathrm{ker} (D)} ( e^{\widehat{\sf Q}^2}) -\mathrm{tr}_{\mathrm{coker} (D)} (e^{\widehat{\sf Q}^2})\,.
\end{align}
Since $\widehat{\sf Q}^2$ generates a linear combination of torus actions,   
$e^{\widehat{\sf Q}^2}$ is thought as a group element.
When the equivariant index is written as a form $\text{ind}(D)=\sum_i n_i e^{w_i}$, where $w_i$ is a weight of a representation of $\widehat{\sf Q}^2$ and $n_i$ is the multiplicity of 
the representation,  the ratio of the functional determinants is obtained by the following rule:  
\begin{align}
\text{ind}(D)=\sum_i n_i e^{w_i} \to \frac{ \mathrm{Det}_{\mathrm{coker}(D) }\widehat{\sf Q}^2}  { \mathrm{Det}_{\mathrm{ker}(D)} \widehat{\sf Q}^2} =\prod_i w_i^{n_i}\,.
\label{eq:rule}
\end{align}

Thus remained task to obtain the one-loop determinant is to evaluate the weights $w_i$ and the multiplicities $n_i$ for the representations of  $ D_{\text{vec}}$ and $D_{\text{hyp}}$.
The expression of $D_{10}$ evaluated in Appendix \ref{ap:diff}  shows that the differential operators $D_{\text{vec}}$ and $D_{\text{hyp}}$  are same type as the ones  appeared in the evaluation
 of the one-loop determinant for 't Hooft loops on $S^1 \times \mathbb{R}^3$ \cite{Ito:2011ea}. 
The  difference  between the differential operators for 't Hooft surface operators here and the ones for  BPS 't Hooft loop operators in \cite{Ito:2011ea} only comes from the spacetime derivative in $\widehat{\sf Q}^2$;
  $\widehat{\sf Q}^2$ in  \eqref{eq:squareQ}   contains the derivative of the  $T^2$-coordinate, whereas 
$\widehat{\sf Q}^2$ for 't Hooft loops  contains the derivative of the  $S^1$-coordinate \cite{Ito:2011ea}. Therefore, for the each Kaluza-Klein (KK) mode along the $T^2$-direction, the index is same as  the equivariant indices $\text{ind} (D_{\text{Bogo}})$  and  $\text{ind} (D_{\text{DH}, \mathcal{R}})$
 evaluated in \cite{Gomis:2011pf, Ito:2011ea}: 
\begin{align}
 \text{ind} (D_{\text{Bogo}, \mathbb{C}})&=-\frac{e^{\pi {\rm i} \epsilon}+e^{-\pi {\rm i} \epsilon} }{2}
\sum_{\alpha \in \text{rt}(\mathfrak{g})}  e^{2 \pi {\rm i} \alpha \cdot {\bm a}} \sum_{k=0}^{|\alpha \cdot {\bm p}| -1} e^{\pi {\rm i} (|\alpha \cdot {\bm p}|-2k-1) \epsilon}\,, \\
  \text{ind} (D_{\text{DH}, \mathcal{R}})&=-\frac{1}{2}\sum_{w \in \Delta(\mathcal{R})}  e^{2 \pi {\rm i} w \cdot {\bm a}}\sum_{k=0}^{|w \cdot {\bm p}| -1} e^{\pi {\rm i} (|w \cdot {\bm p}|-2k-1) \epsilon}\,,
\end{align}
Here we choose a saddle point specified by a magnetic charge ${\bm p}$.
By summing over the KK modes along $T^2$ we obtain the indices of $D_{\text{vec}}$ and $D_{\text{hyp}}$ 
on $T^2 \times \mathbb{R}^3$:
\begin{align}
\mathrm{ind}(D_{\text{vec}})&= \sum_{m,n \in \mathbb{Z}} e^{2 \pi {\rm i} (m+n \tau)} \text{ind} (D_{\text{Bogo}, \mathbb{C}}) \,,
\label{eq:Gindex1}
  \\
\mathrm{ind}(D_{\text{hyp}})&= -\sum_{m,n \in \mathbb{Z}} e^{2 \pi {\rm i} (m+n \tau)} 
\sum_{f=1}^{N_F} \left( e^{- 2 \pi {\rm i} m_f }\text{ind} (D_{\text{DH},\mathcal{R}}) +e^{ 2 \pi {\rm i} m_f }\text{ind} (D_{\text{DH},\mathcal{R}})|_{{\bm a} \to -{\bm a}}\right) \,.
\label{eq:Gindex2}
\end{align}
Note that if the KK momentum contributions $\sum_{m,n \in \mathbb{Z}} e^{2 \pi {\rm i} (m+n \tau)}$ for $T^2$ in \eqref{eq:Gindex1} and \eqref{eq:Gindex2} are replaced by  $\sum_{m \in \mathbb{Z}} e^{2 \pi {\rm i} m}$ for $S^1$,
 we obtain the equivariant indices for the one-loop determinants of the BPS  't Hooft loop operators on $S^1 \times \mathbb{R}^3$.
By using the relation \eqref{eq:rule} with
 \eqref{eq:Gindex1} and \eqref{eq:Gindex2}, up to overall normalization constants, 
we obtain the non-regularized expression of the one-loop determinants around a saddle point 
specified by a magnetic charge ${\bm p}$:
{\footnotesize{
\begin{align}
&Z^{\text{5d.vec}}_{1\mathchar `-\text{loop}}({\bm a},\bm{p}, \epsilon, \tau)= 
 \left[\prod_{\alpha \in \mathrm{rt}(\mathfrak{g}) } \prod_{k=0}^{|\alpha \cdot \bm{p}|-1} 
\prod_{m, n\in \mathbb{Z}} (m+n \tau+ \alpha \cdot {\bm a} + \left( \frac{|\alpha \cdot \bm{p}| }{2}-k \right) \epsilon
 \right] ^{-\frac{1}{2}},
\label{eq:1loopvec2}
\\
&Z^{\text{5d.hyp}}_{1 \mathchar `-\text{loop}}({\bm a}, {\bm m}, \bm{p}, \epsilon, \tau)=  
   \left[\prod_{{ w} \in \Delta(\mathcal{R}) } \prod_{f=1}^{N_F} \prod_{k=0}^{|{w} \cdot \bm{p}|-1} \prod_{m, n\in \mathbb{Z}}   \left (m+n \tau+ { w} \cdot {\bm a} -{ m}_f + \left( \frac{|{ w} \cdot \bm{p}| -1}{2}-k \right) \epsilon \right) \right]^{\frac{1}{2}}, 
\label{eq:1loophy2}
\end{align}
}}
By using the zeta function regularization, we obtain the one-loop determinants  \eqref{eq:1loopvec}, \eqref{eq:1loophy}.

%%%%%%%%%%%%%%%%%%%%%%%%%%%%%%%%%%%%%%%%%%%%%%%%%%%%%%%%%
\section{Monopole bubbling effects via branes and elliptic genera}
\label{sec:bubbling}
For BPS 't Hooft loops in 4d $\mathcal{N}=2$ $U(N)$ and $SU(N)$ gauge theories on $S^1 \times \mathbb{R}^3$, 
monopole bubbling effects are  given by Witten indices for supersymmetric quantum mechanics (SQM) arising from the low energy world volume theories on 
D1-branes \cite{Brennan:2018rcn, Brennan:2018yuj, Assel:2019iae} in the type IIB string theory. 
Monopole bubbling effects were  also  studied in \cite{Hayashi:2020ofu} in terms of the type IIA string theory. 
By taking a T-dual of the brane configuration of \cite{Hayashi:2020ofu} in the $x^4$-direction, 
we obtain  monopole bubbling effects $Z_{\text{mono}}$ for the surface operators on $T^2 \times \mathbb{R}^3$ which are given  by elliptic genera for 2d gauged linear sigma models (GLSMs) living on $T^2$.

%%%%%%%%%%%%%%%%%%%%%
\subsection{Brane construction for t' Hooft surface operators and monopole bubbling}
\label{sec:section41}
\begin{table}[t]
\begin{center}
\begin{tabular}{cccccc|cc|ccc}
&0&1&2&3&4&5&6&7&8&9
\\
\hline
D5&$\times$&$\times$&$\times$&$\times$&$\times$&&$\times$ 
\\
NS5 &$\times$&$\times$&$\times$&$\times$&$\times$&$\times$&&
\\
\hline
D3 &$\times$&&&&$\times$&$\times$&$\times$
\\
NS5'&$\times$&&&&$\times$&&$\times$&$\times$&$\times$&$\times$
\end{tabular}
\end{center}
\caption{The symbol $\times$ denotes the directions in which D-branes and NS5-branes extend. }
\label{table:brane}
\end{table}

\begin{figure}[thb]
\centering
\subfigure[]{\label{fig:diracmono1}
\includegraphics[width=5cm]{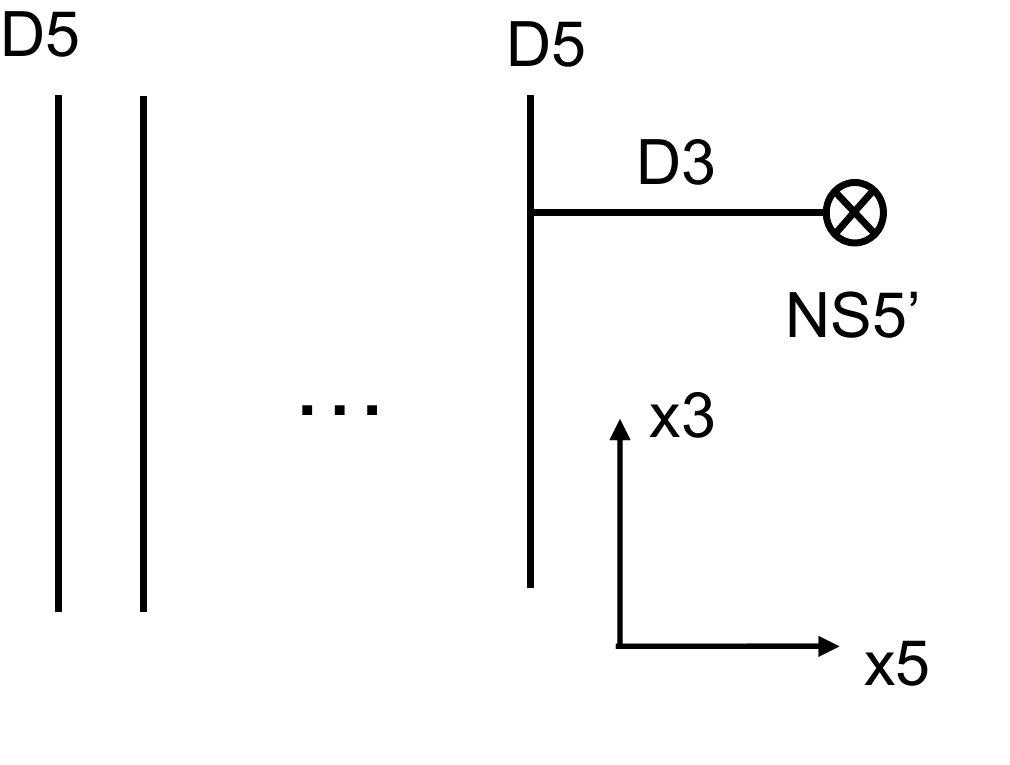}}
\hspace{1cm}
\subfigure[]{\label{fig:diracmono2}
\includegraphics[width=5cm]{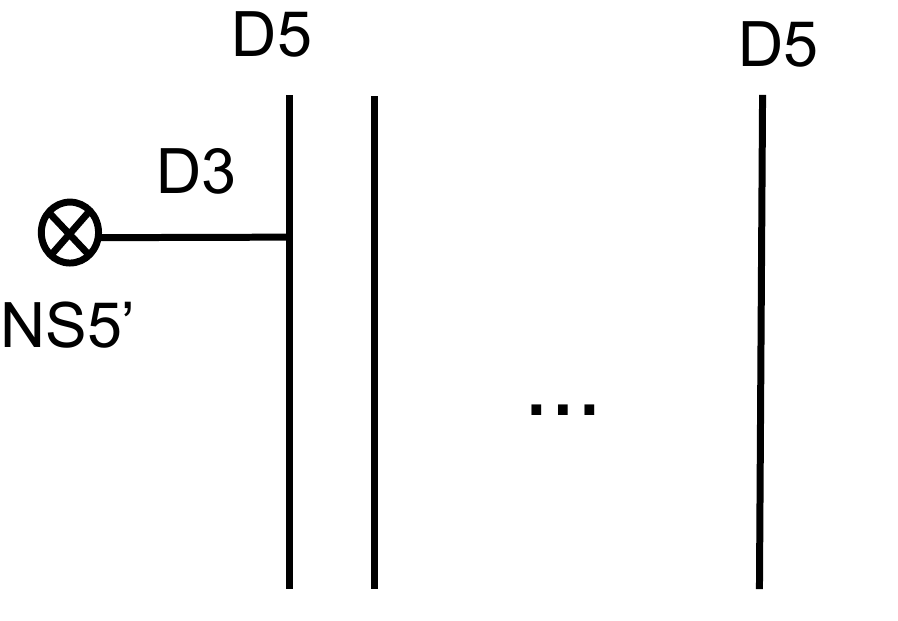}}
\subfigure[]{\label{fig:antisym1}
\includegraphics[width=5cm]{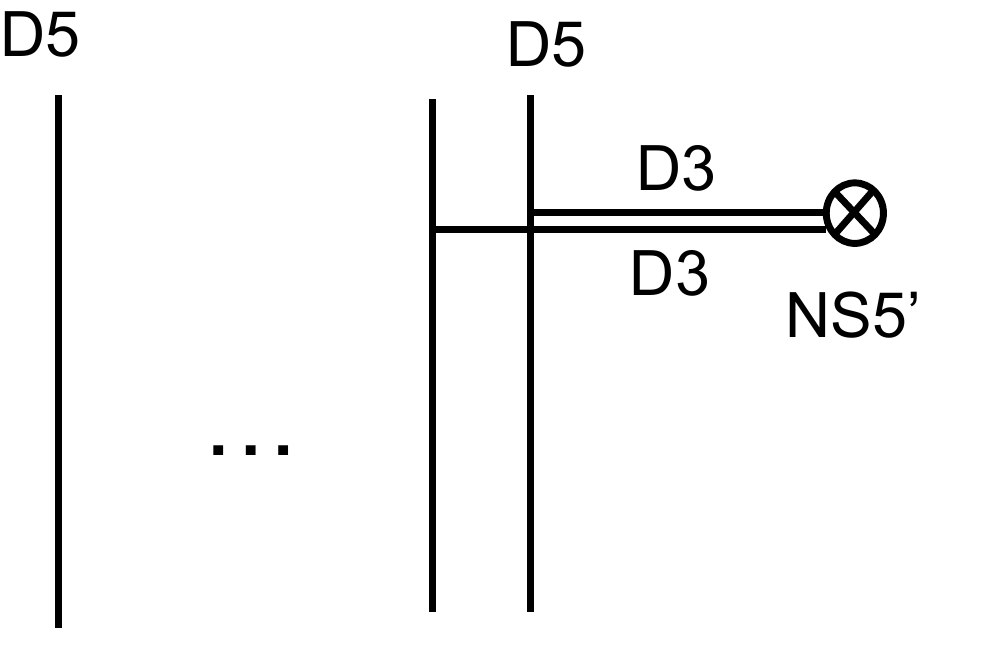}}
\hspace{1cm}
\subfigure[]{\label{fig:symmono1}
\includegraphics[width=5cm]{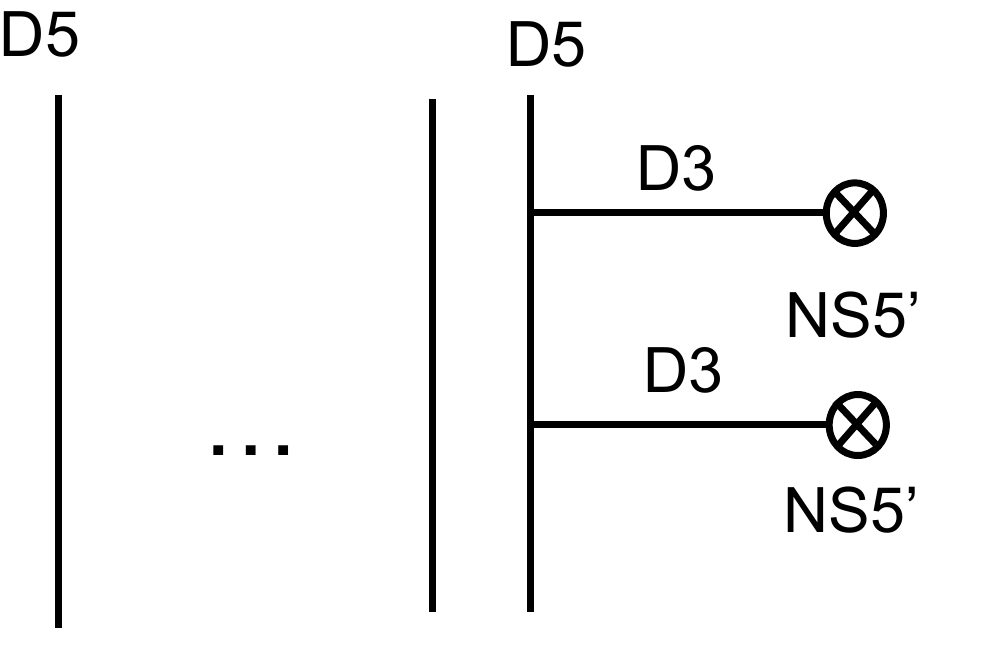}}
\caption{(a): A brane configuration for  a magnetic charge ${\bm B}={\bm e}_1$. (b):  A brane configuration for  a magnetic charge ${\bm B}=-{\bm e}_N$.
(c): A brane configuration for  ${\bm B}={\bm e}_1+{\bm e}_2$. (d): A brane configuration for a magnetic charge ${\bm B}=2 {\bm e}_1$. 
  }
\label{fig:diracmono}
\end{figure}

%\rem{
\begin{figure}[thb]
\centering
\subfigure[]{\label{fig:adjop1}
\includegraphics[width=5cm]{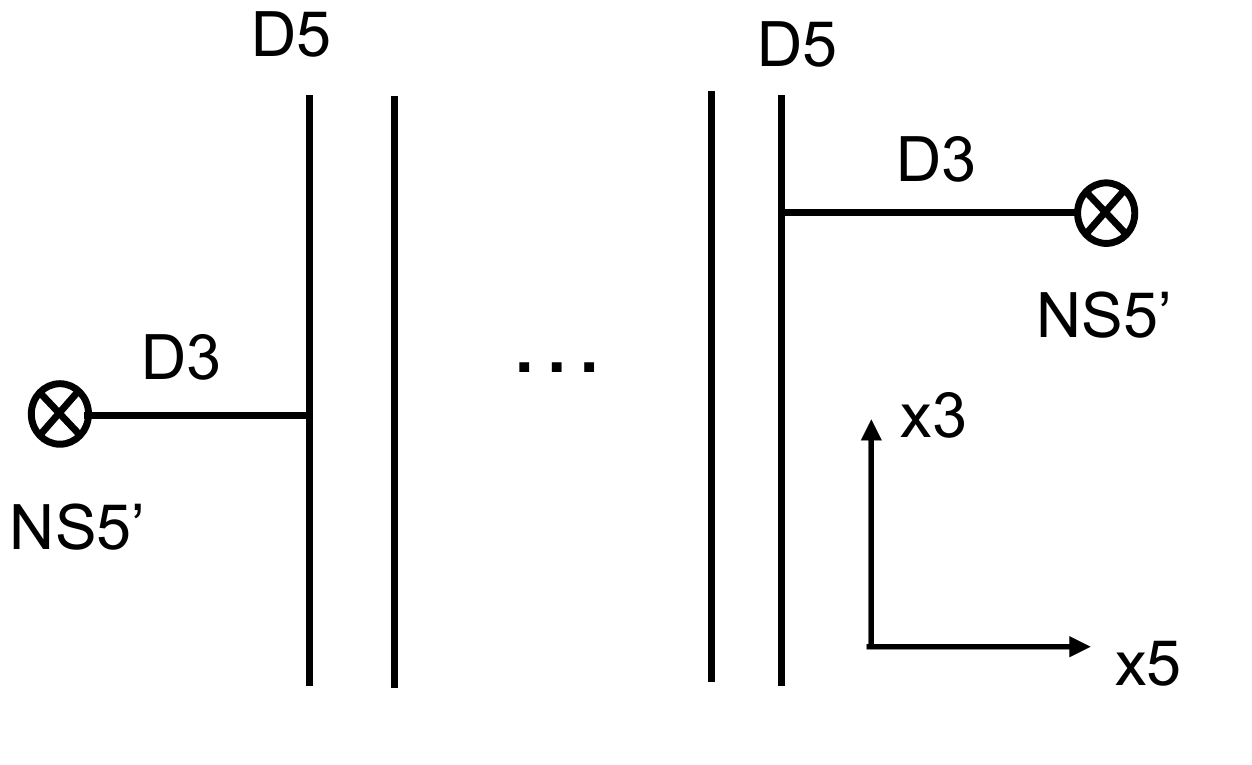}}
\hspace{1cm}
\subfigure[]{\label{fig:adjop2}
\includegraphics[width=5cm]{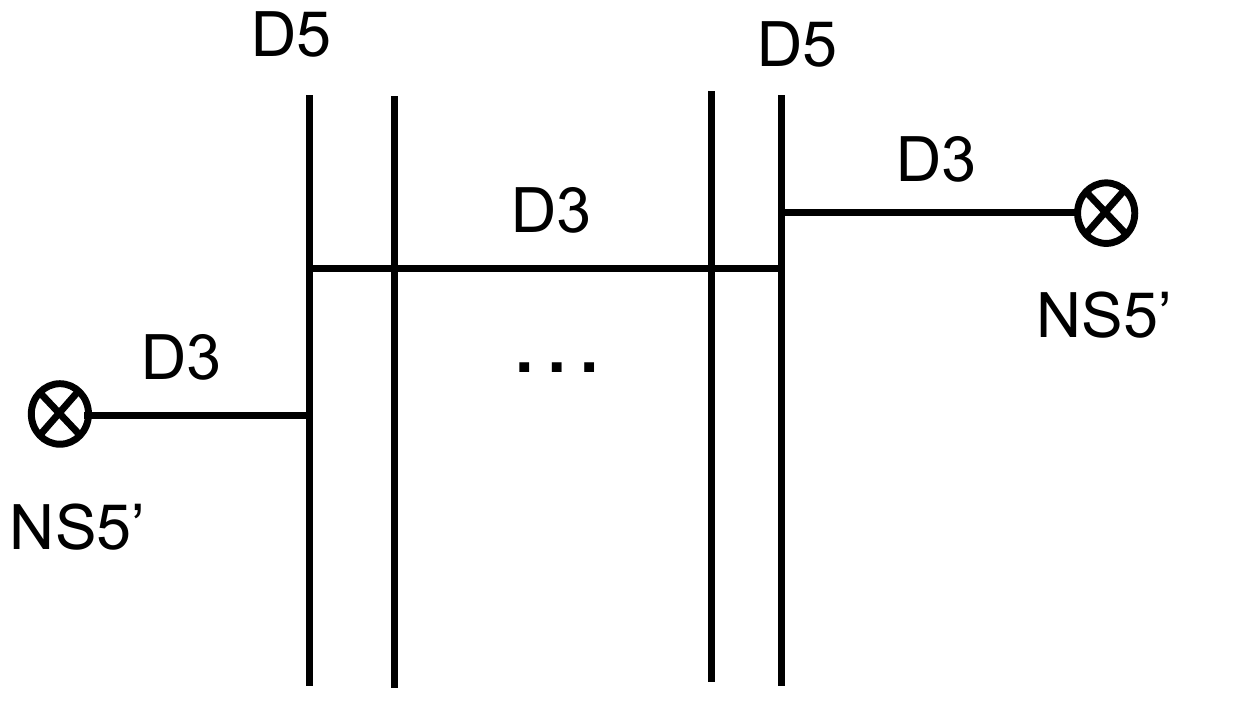}}
\subfigure[]{\label{fig:adjop3}
\includegraphics[width=5cm]{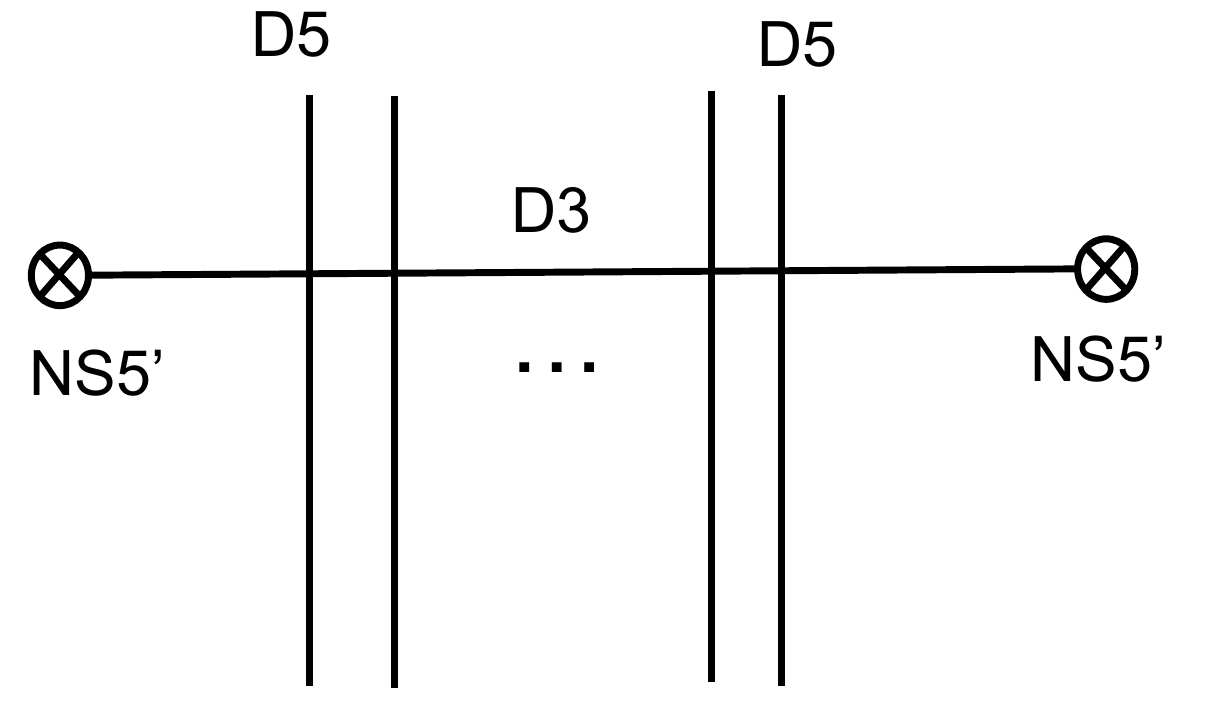}}
\hspace{1cm}
\subfigure[]{\label{fig:quiver1}
\includegraphics[width=1cm]{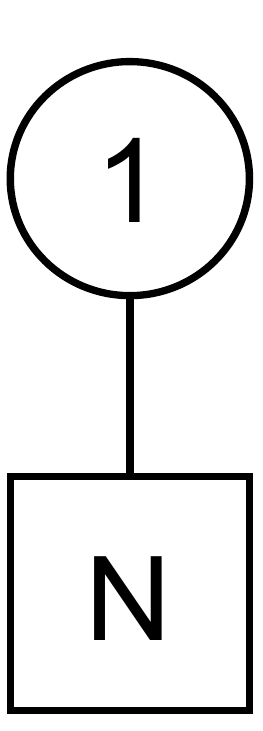}}
\caption{(a): The brane configuration for the surface operator with the magnetic charge $ {\bm B}={\bm e}_1 - {\bm e}_N$. 
 (b) A D3-brane suspended between the leftmost D5-brane and the rightmost D5-brane is introduced to describe the 't Hooft-Polyakov monopole with the charge ${\bm B}=-{\bm e}_1 + {\bm e}_N$. 
(c): When  the  positions of the three  D3-branes in the $x^3$-direction   coincide,  the D3-branes  reconnect and form a single D3-brane. Then the charge is reduced to ${\bm p}=0$ which is  the brane interpretation of 
the monopole bubbling effect.
 (d): The quiver diagram for the world volume theory on
the D3-brane in Figure (c). The circle with $1$ denotes 2d $\mathcal{N}=(4,4)$ $U(1)$ vector multiplet. The solid line connected between the circle and the box with $N$ denotes the $N$-tuple 2d $\mathcal{N}=(4,4)$ hypermultiplets with the gauge charge $+1$.   
  }
\label{fig:screen1}
\end{figure}
First we explain a 5d $\mathcal{N}=1$ $U(N)$ gauge theory arising from the low energy world volume theory on D5-NS5-brane system.
The brane configuration is specified by Figure \ref{table:brane}. 
When $N$ D5-branes are suspended between two NS5-branes in the $x^6$-direction,  
a 5d    $\mathcal{N}=1$ $U(N)$ vector multiplet arises from the low energy world volume theory on the D5-branes.  
When $N$ D5-branes and $N^{\prime}$ D5-branes are attached to an NS5-brane from the left side and the right side in the $x^6$-direction, respectively, 
a 5d  $\mathcal{N}=1$  hypermultiplet in the $U(N) \times U(N^{\prime})$ bi-fundamental representation arises from the open strings ended on the $N$ D5-branes and the $N^{\prime}$ D5-branes separated by the NS5-brane.
If these $N$ (resp. $N^{\prime}$) D5-branes are suspended between two NS5-branes, $U(N)$ (resp. $U(N^{\prime})$) is a  gauge group. On the other hand, if 
 the $N$ (resp. $N^{\prime}$) D5-branes are semi-infinite in the $x^6$-direction,  $U(N)$ (resp. $U(N^{\prime})$) is a flavor symmetry group. 
If the $x^{6}$-direction is compactified by a circle, we obtain circular quiver gauge theories with $U(N)$ gauge groups.
In this paper, we compute explicitly monopole bubbling effects for   5d $U(N)$ $\mathcal{N}=1^{*}$ gauge theories. The $\mathcal{N}=1^*$ $U(N)$ gauge theory is obtained by the $N$ D5-branes are   ended on a single NS5-brane and $x^{6}$-direction is compactified by the circle. 
Monopole bubbling effects for other gauge theories will be studied  in elsewhere.

An 't Hooft  surface operator  is described by D3-branes  stretched between D5-branes and  NS5'-branes  in the $x^5$-direction. 
  Here we write an NS5-brane extending in the $x^{0,4,6,7,8,9}$-directions as an NS5'-brane to distinguish it from 
an NS5-brane extending in the $x^{0,1,2,3,4,5}$-directions.
For later convenience, we call $N$ D5-branes as the first, the second, ..., the $N$-th D5-brane from the right to the left in the $x^5$-direction. The magnetic  charge ${\bm B}$ is read off from brane configurations as
\begin{itemize}
\item A D3-brane attached to the $i$-th D5-brane from the right side  corresponds to  a magnetic charge ${\bm B}={\bm  e}_i$.
\item A D3-brane attached to the $i$-th D5-brane from the left side  corresponds to  a magnetic charge ${\bm B}=-{ \bm e}_i$.
\end{itemize}
Here ${\bm e}_i=\mathrm{diag}(0, \cdots,0,\stackrel{i}{1},0\cdots,0)$.
Examples of brane configurations for surface operators and their magnetic charges are listed in Figure \ref{fig:diracmono}.

To describe monopole bubbling effects in the brane set up, we introduce 
a D3-brane suspended between the $i$-th D5-brane and the $j$-th D5-brane which is interpreted as a smooth 't Hooft-Polyakov monopole
 with ${\bm B}={ \bm e}_i-{\bm e}_j$ \cite{Diaconescu:1996rk}, see Figure \ref{fig:adjop2}. 
When the $x^3$-coordinate of the D3-branes for the surface operators and the one of a D3-brane for the  't Hooft-Polyakov monopole
coincide, the magnetic charge is screened and monopole bubbling occurs, see  \ref{fig:adjop3}. The low energy world volume theory on D3-branes suspended between 
  NS5'-branes is 2d supersymmetric GLSM. This T-dual picture of the proposal in \cite{Brennan:2018yuj, Hayashi:2020ofu} suggests that
the  supersymmetric indices, i.e., elliptic genera for the world volume theories of D3-branes ended on NS5'-branes give
the monopole bubbling effects $Z_{\text{mono}}$. 

%%%%%%%%%%%%%%%%%%%%%%%%%%%%%%%%%%%%%%%%%
\subsection{Elliptic genera  for monopole bubbling effects}
By using the localization formula for elliptic genera in \cite{Benini:2013nda, Benini:2013xpa}, we can express the elliptic genus for the monopole bubbling effects  in terms of  residue integrals called Jeffrey-Kirwan (JK) residues:
\begin{align}
 Z^{(\eta)}_{\text{mono}} ({\bm a}, {\bm m}, {\bm p}, {\bm B}, \epsilon, \tau) &=\frac{1}{|W_{G_{\text{2d}}}|} \sum_{u_{\ast} \in \mathfrak{M}_{\text{sing}}  }\mathop{\text{JK-Res}}_{ u = {u}_*} ({Q}_*,{\eta}) \nonumber \\
& \quad \times \prod Z_{\text{2d.vec}} \prod  Z_{\text{2d.hyp}} 
%\prod    Z_{\text{2d.fermi}} 
\, d u^1 \wedge \cdots \wedge  d u^{\mathrm{rk}(\mathfrak{g}_{\text{2d}})}\,.
\label{eq:Zmono}
\end{align}
 Here  $G_{\text{2d}}$ and $\mathfrak{g}_{\text{2d}}$ are the gauge group and its Lie algebra of 2d GLSM, respectively.  $\text{rk} (\mathfrak{g})$ is the rank of the Lie algebra $\mathfrak{g}$. 

For 5d $\mathcal{N}=1^*$ $U(N)$ gauge theory, the GLSMs for the monopole bubbling  preserve the 2d $\mathcal{N}=(4,4)$ supersymmetry ( more precisely the mass deformation of  $\mathcal{N}=(4,4)$ supersymmetry).
For $G_{\text{2d}}=U(n)$ 2d $\mathcal{N}=(4,4)$ GLSM, the one-loop determinant for the vector multiplet is 
\begin{align}
Z_{\text{2d.vec}} =
\left( 2 \pi \eta(\tau)^3 \right)^{n}
   \prod_{1 \le a \neq b \le n} \vartheta_1  ( u_a-u_b ) \cdot \prod_{ 
a,b=1 }^n  \frac{  \vartheta_1( u_a-u_b+\epsilon) } 
{  \vartheta_1 (u_a-u_b \pm {m}_{\text{ad}}  + \frac{1}{2}\epsilon)}   \,.
\label{one-loop-2d-vec} 
\end{align}
A fugacity $m_\text{ad}$ corresponds to  the flavor fugacity  for  5d $\mathcal{N}=1^*$ theory. 
The one-loop determinant for a 2d $\mathcal{N}=(4, 4)$ hypermultiplet in the bifundamental representation of  $U(n) \times U(n^{\prime})$ is 
\begin{align}
&Z_{\text{2d.hyp}}
= \prod_{a =1}^n \prod_{b=1}^{n^{\prime}}  \frac{\vartheta_1 (\pm ( u_a -u^{\prime}_b)-m_{\text{ad}})}
{\vartheta_1( \pm ( u_a -u^{\prime}_b) +\frac{1}{2}  \epsilon) }  \,.
\label{one-loop-2d-hyp}
\end{align}
When $U(n')$ is a flavor symmetry of 2d GLSM, fugacity $u^{\prime}$ corresponds to components of  the flat connection ${\bm a}$ of the five-dimensional gauge field.

$\mathop{\text{JK-Res}}_{ u = {u}_*} ({Q}_*,{\eta})$ in \eqref{eq:Zmono} denotes the Jeffrey--Kirwan residue at a point $u_*$ defined as follows. 
We consider a situation that  $\mathrm{rk}(\mathfrak{g}_{\text{2d}})$ hyperplanes of  codimension one, called singular hyperplanes ${Q}_i \cdot  ({u}-{u}_*) =\sum_{a=1}^{\mathrm{rk}(\mathfrak{g}_{\text{2d}})} {Q}^a_i({u}^a-{u}^a_*)=0$
 $( i=1, \cdots, \mathrm{rk}(\mathfrak{g}_{\text{2d}}) )$  
intersect at a point $u_*=(u^1_*, \cdots, u^{\mathrm{rk}(\mathfrak{g}_{\text{2d}})}_*)$ in the $u$-space.   
In our case, $Q_i=(Q^1_i, \cdots, Q^{\text{rk}(\mathfrak{g}_{\text{2d}})}_i) \in \mathbb{R}^{\mathrm{rk}(\mathfrak{g}_{\text{2d}})}$ 
 is a weight vector appearing in the denominators of the one-loop determinants \eqref{one-loop-2d-vec} and \eqref{one-loop-2d-hyp}. 
Then the JK residue at the point $u_*$ is  defined by
\begin{equation}\label{JKnonde}
\begin{aligned}
&\mathop{\text{JK-Res}}_{ u = {u}_*} ({Q}_*,{\eta}) \frac{du^1 \wedge\cdots\wedge du^{\mathrm{rk}(\mathfrak{g}_{\text{2d}})} }{{Q}_1 \cdot ({u}-{u}_*)\cdots Q_{\mathrm{rk}(\mathfrak{g}_{\text{2d}})} \cdot  ({u}-{u}_*) }
\\
& \qquad \qquad \qquad \qquad
=
\left\{
\begin{array}{cl}
\frac{1}{|\det({Q}_1,\ldots,{Q}_{\text{rk}(\mathfrak{g}_{\text{2d}})})|} & \text{ if } {\eta}\in \text{Cone}({Q}_1,\ldots,{Q}_{\text{rk}(\mathfrak{g}_{\text{2d}})})\,, \\
0 & \text{ otherwise}\,.
\end{array}
\right. 
\end{aligned}
\end{equation}
Here  $\text{Cone}({Q}_1,\ldots,{Q}_{\text{rk}(\mathfrak{g}_{\text{2d}})})= \sum _{i=1}^{\text{rk}(\mathfrak{g}_{\text{2d}})} \mathbb{R}_{> 0} Q_i$ and each $Q_i$ is a $\text{rk}(\mathfrak{g}_{\text{2d}})$-dimensional vector. 
 The sum $\sum_{u_{\ast}  }$ runs over all the points $u_{\ast}$, 
where  $N^{\prime}$ with $N^{\prime} = \text{rk}(\mathfrak{g}_{\text{2d}})$ singular hyperplanes meet at a point and  the condition ${\eta}\in \text{Cone}({Q}_1,\ldots,{Q}_{\text{rk}(\mathfrak{g}_{\text{2d}})})$ is satisfied.
 If $N^{\prime}$ singular hyperplanes with $N^{\prime} > \text{rk}(\mathfrak{g}_{\text{2d}})$ intersect at a point, we apply the constructive definition  of  the JK residue in \cite{Benini:2013xpa}. All the examples treated in Section \ref{sec:section6} satisfy the condition $N^{\prime} = \text{rk}(\mathfrak{g}_{\text{2d}})$.

%%%%%%%%%%%%%%%%%%%%%%%%%%%%%%%%%%%%%%%%
\section{Deformation quantization and elliptic Ruijsenaars operators }
\label{sec:section5}
In \cite{Okuda:2019emk}  we have shown that  the deformation quantization of 
 SUSY localization formula of the 't Hooft loops  in the anti-symmetric representation  in 4d $\mathcal{N}=2^{*}$ $U(N)$ gauge theory  can be 
identified with trigonometric Macdonald operators by redefinitions of variables. We also gave the  identification between   SUSY localization formulas for monopole operators
 and representation of monopole operators as    difference  operators of rational type in \cite{Braverman:2016aa}. These observations motivate us to 
relate  a deformation quantization of vevs of 't Hooft surface operators to difference operator of elliptic type.

We  regard ${\bm a}=(a_1, \cdots, a_{\mathrm{rk}(\mathfrak{g}) })$, ${\bm b}=(b_1, \cdots, b_{\mathrm{rk}(\mathfrak{g}) })$   as the coordinate and momentum variables, respectively, and define the quantization   by
\begin{align}
[\hat{a}_i, \hat{a}_j]=0, \quad [\hat{b}_i, \hat{b}_j]=0, \quad 
[\hat{b}_i, \hat{a}_i]=\epsilon C_{i j}\,.
\end{align}
where, and $C_{i j}$ is the $(i,j)$-component of the inverse matrix of the Killing form $C^{i j}=\mathrm{Tr}(H_i H_j)$, where  $\{ H_i \}_{i=1}^{\mathrm{rk}(\mathfrak{g})}$ is bases of the Cartan subalgebra  of $\mathfrak{g}$.\footnote{ For the simple Lie algebras, we normalize   $H_i$  so that $C^{ij}$ agrees with the Cartan matrix. For $U(N)$ gauge theory, $C^{i j}$ is normalized as $C^{i j}=\delta^{i j}$.}

We define the   quantization of a function $f({\bm a}, {\bm b})$   by the Weyl-Wigner transform:
\begin{align}
\widehat{f}(\widehat{\bm a}, \widehat{\bm b})
: =\left[
\exp \left( \frac{\epsilon}{2}  \sum_{i,j=1}^N C_{i j} \partial_{{b}_i} \partial_{{a}_j} \right) f ({\bm a},{\bm b})\right]\Bigg |_{{\bm a} \mapsto \widehat{\bm a}, {\bm b}\mapsto \widehat{\bm b}}.
\label{eq:WeylWigner}
\end{align}
After the derivative is taken in \eqref{eq:WeylWigner}, the ordering of $\hat{a}_i$ and $\hat{b}_i$ is defined as follows.  All the $a_i$'s should be on left side and all $b_i$'s shoud be on the right side.
Next we define the Moyal product  $f * g$ of $f$ and $g$  by
\begin{align}
f({\bm a}, {\bm b})*
g({\bm a}, {\bm b}):=
\exp \left( \frac{\epsilon}{2} \sum_{i,j =1}^{N} C^{i j}  (\partial_{\underline{a}_i} \partial_{b_i} -\partial_{a_i} \partial_{\underline{b}_i} ) \right) f({\bm a}, {\bm b}) g(\underline{\bm a}, \underline{\bm b}) |_{\underline{\bm a}={\bm a}, \underline{\bm b}={\bm b}}\,.
\label{eq:Moyalp}
\end{align}
Since  the Weyl-Wigner transform satisfies the following relation 
\begin{align}
\widehat{f*
g}= \hat{f} \hat{g},
\end{align}
the algebra of the deformation quantization of surface operators can be studied in term of 
the Moyal products.
For example, the commutativity of the operators $\hat{f}$ and $\hat{g}$ is equivalent to that of 
$f$ and $g$ with respect to the Moyal product:
\begin{align}
[ \hat{f}, \hat{g}] = 0 \leftrightarrow f * g -g * f = 0.
\end{align}
In particular, we will see that the commutativity of  elliptic Ruijsenaars operators is rephrased as the absence of  wall-crossing phenomena ($\eta$-independence)  of elliptic genera for the monopole bubbling effects. 
 Recently the author of \cite{Maruyoshi:2020cwy} observed that the localization formula \cite{Ito:2011ea} of 't Hooft  loop operators  with  particular charges in 4d $\mathcal{N}=2^*$  gauge theory agrees with 
a trigonometric limit of  type-$A$  elliptic Ruijsenaars operators  \cite{Ruijsenaars:1986pp}. 
Since the  KK modes along the $T^2$-direction give an elliptic deformation of the localization formula of the 't Hooft loops,  we expect the deformation quantization  of 't Hooft surface operator itself agrees  with elliptic Ruisenaars operators.  
In fact, we show that the deformation quantization of 't Hooft  surface operators with specific magnetic charge in 5d $\mathcal{N}=1^*$ theory is 
identified with  elliptic Ruijsenaars operators. We also study the 
algebra of 't Hooft surface operators defined by  the Moyal product .

From the the localization formula \eqref{eq:localizationfm},  the vevs of  't Hooft surface operators $S_{\bm B}$ with ${\bm B}=\sum_{k=1}^{\ell} {\bm e}_k$ and ${\bm B}=-\sum_{k=\ell+1}^{N} {\bm e}_k$
 in the $\mathcal{N}=1^*$ $U(N)$ gauge theory are given by\footnote{A dominant coweight ${\bm B}$ of $U(N)$ is $\text{diag}(B_1,B_2,\cdots, B_N)$ with $B_1 \ge B_2 \ge \cdots \ge B_N$ and $B_i \in \mathbb{Z}$.}
\begin{align}
\langle S_{\sum_{k=1}^{\ell} {\bm e}_k} \rangle &=
\sum_{I \subset \{1, \cdots, N \} \atop |I|=\ell}  \left( \prod_{i \in I , j \notin I} 
\frac{ \vartheta_1 \left( a_{i} -a_{ j}-m_{\mathrm{ad}} \right)  \vartheta_1  \left( a_{j} -a_{ i} -m_{\mathrm{ad}} \right)}{\vartheta_1 \left( a_{i} -a_{ j}+\frac{\epsilon}{2} \right)  \vartheta_1 \left( a_j -a_i+\frac{\epsilon}{2} \right) } \right)^{\frac{1}{2}} \prod_{i \in I}  e^{  b_i}, 
 \\
\langle S_{-\sum_{k=\ell+1}^{N} {\bm e}_k} \rangle&=\sum_{I \subset \{1, \cdots, N \} \atop |I|=\ell}  \left( \prod_{i \in I , j \notin I} 
\frac{ \vartheta_1 \left( a_{i} -a_{ j}-m_{\mathrm{ad}} \right)  \vartheta_1  \left( a_{j} -a_{ i} -m_{\mathrm{ad}} \right)}{\vartheta_1 \left( a_{i} -a_{ j}+\frac{\epsilon}{2} \right)  \vartheta_1 \left( a_j -a_i+\frac{\epsilon}{2} \right) } \right)^{\frac{1}{2}} \prod_{i \in I}  e^{  -b_i}
\end{align}
Here $m_{\mathrm{ad}}$ is the flavor fugacity for the adjoint hypermultiplet in five dimensions.
$I$ is a subset of $\{1,\cdots, N \}$ with the cardinality $|I|=\ell$.
%%%%%%%%%%%%%%%%%%%%%%%%%%%
\rem{
where we define
\begin{align}
{\bm B}^{(\pm)}_{\ell}=\pm\mathrm{diag}(\underbrace{1,\cdots,1}_{\ell},0,\cdots,0)\pm\frac{\ell}{N} \mathrm{diag}(1,\cdots,1)
\end{align}
and
\begin{align}
&a_i=a^{\prime}_i-a^{\prime}_{i-1}, a^{\prime}_0=a^{\prime}_N=0, \text{ for }  i=1,\cdots,N\,, \\
&b_i=b^{\prime}_i-b^{\prime}_{i-1},  b^{\prime}_0=b^{\prime}_N=0 , \text{ for }  i=1,\cdots,N\,.
\end{align}
}
%%%%%%%%%%%%%%%%%%%%%%%%%%%
Then the Weyl-Wigner transform of the vevs of surface operators  are written as 
\begin{align}
 \widehat{S}_{\sum_{k=1}^{\ell} {\bm e}_k}  &=
\sum_{I \subset \{1, \cdots, N \} \atop |I|=\ell}  \left( \prod_{i \in I , j \notin I} 
\frac{ \vartheta_1 \left( a_{i} -a_{ j}-m_{\mathrm{ad}}+\frac{\epsilon}{2} \right)  \vartheta_1  \left( a_{j} -a_{ i} -m_{\mathrm{ad}}-\frac{\epsilon}{2} \right)}
{\vartheta_1 \left( a_{i} -a_{ j}+\epsilon \right)  \vartheta_1 \left( a_j -a_i \right) } \right)^{\frac{1}{2}} \prod_{i \in I}  e^{ \epsilon  \frac{\partial}{\partial a_{i}} }\,,
\label{eq:defsur1} 
 \\
\widehat{S}_{-\sum_{k=\ell+1}^{N} {\bm e}_k} &=\sum_{I \subset \{1, \cdots, N \} \atop |I|=\ell}  \left( \prod_{i \in I , j \notin I} 
\frac{ \vartheta_1 \left( a_{i} -a_{ j}-m_{\mathrm{ad}}-\frac{\epsilon}{2} \right)  \vartheta_1  \left( a_{j} -a_{ i} -m_{\mathrm{ad}} +\frac{\epsilon}{2}\right)}{\vartheta_1 \left( {a}_{i} -{a}_{ j} \right)  \vartheta_1 \left( a_j -a_i+\epsilon \right) } \right)^{\frac{1}{2}} \prod_{i \in I}  e^{-  \epsilon  \frac{\partial}{\partial a_{i}} }\,.
\label{eq:defsur2}
\end{align}
Here we use a representation $\hat{a}_i=a_i$ and $\hat{b}_i =\epsilon \frac{\partial}{\partial a_i}$, and 
find that the deformation quantization of the 't Hooft  surface operators  \eqref{eq:defsur1} and \eqref{eq:defsur2} agree with  type-$A$ elliptic Ruijsenaars operators \cite{Ruijsenaars:1986pp}.

%%%%%%%%%%%%%%%%%%%%%%%%%%%%%%%%%%%%%%%%
\section{Products of 't Hooft surface operators and monopole bubbling}
\label{sec:section6}
In this section, we evaluate  explicitly elliptic genera for monopole bubbling effects  with small values of magnetic charges and compare them with  results from the Moyal product.
We consider monopole bubbling effects in the monopole surface operator $S_{\bm B }$ for ${\bm B}=(1,0, \cdots, 0, -1)=\mathrm{diag}(1,0, \cdots, 0, -1)$ and ${\bm B}=(1,1,0,\cdots, 0, -1,-1)=\mathrm{diag} (1,1,0,\cdots, 0, -1,-1)$ in  $\mathcal{N}=1^*$ $U(N)$ gauge theory.   

%%%%%%%%%%%%%%%%%%%%%%%%%%%%%%%%%%%%%%%%%
\subsection{Surface operator $S_{(1,0,\cdots, 0,-1)}$}
From the localization formula \eqref{eq:localizationfm} and the definition of the Moyal product \eqref{eq:Moyalp}, two different orderings of 
 the Moyal product of  $\langle S_{{\bm e}_1} \rangle$ and $\langle S_{-{\bm e}_N} \rangle$ are evaluated as
\begin{align}
\langle S_{{\bm e}_1}  \rangle *\langle S_{-{\bm e}_N} \rangle
&= \sum_{1 \le i \neq j \le N}   e^{  { b_i -b_j}    }Z^{\text{5d}}_{1\text{-loop}} ({\bm p}={\bm e}_i-{\bm e}_j) \nonumber  \\
&+\sum_{i=1}^N \prod_{j=1 \atop j \neq i}^{N}  \frac{\vartheta_1 (  a_i -a_j-m_{\text{ad}}+\frac{1}{2} \epsilon   ) \vartheta_1 (  a_j -a_i-m_{\text{ad}}-  \frac{1}{2} \epsilon  ) }
{\vartheta_1(  a_i -a_j+  \epsilon) \vartheta_1(  a_j -a_i ) }\,, 
\label{eq:Moyaladj1}
\\
\langle S_{-{\bm e}_N} \rangle *
\langle S_{ {\bm e}_1 } \rangle&=\sum_{1 \le i \neq j \le N}   e^{  { b_i -b_j}    }Z^{\text{5d}}_{1\text{-loop}} ({\bm p}={\bm e}_i-{\bm e}_j) \nonumber  \\
&+\sum_{i=1}^N \prod_{j=1 \atop j \neq i}^{N}  \frac{  \vartheta_1 (  a_i -a_j-m_{\text{ad}}-\frac{1}{2} \epsilon   ) \vartheta_1 (  a_j -a_i-m_{\text{ad}}+  \frac{1}{2} \epsilon  )}
{\vartheta_1(  a_i -a_j ) \vartheta_1(  a_j -a_i+  \epsilon)}
\,. 
\label{eq:Moyaladj2}
\end{align}
Here the explicit form of the  one-loop determinant $Z^{\text{5d}}_{1\text{-loop}} ({\bm p}={\bm e}_i-{\bm e}_j)$ is  summarized in Appendix \ref{App:1loop}.
Although the second line in \eqref{eq:Moyaladj1} looks different from the one in  \eqref{eq:Moyaladj2}, we will show these terms are actually same by using the contour integral expression of an elliptic genus.  
This leads to the Moyal products of $\langle S_{-{\bm e}_N} \rangle$ and $\langle S_{  {\bm e}_1  } \rangle$ commute each other. 
$\langle S_{{\bm e}_1} \rangle *
\langle S_{-{\bm e}_N} \rangle 
 =
\langle S_{{\bm e}_1} \rangle *
\langle S_{ -{\bm e}_N } \rangle
$.

We evaluate the monopole bubbling effect in the expectation value of $S_{\bm B}$ with ${\bm B}={\bm e}_1 - {\bm e}_N=(1,0\cdots, 0,-1)$ and 
study the algebra of 't Hooft surface operators.  
From the localization formula \eqref{eq:localizationfm},   $\langle S_{{\bm e}_1 - {\bm e}_N} \rangle$ is given by
\begin{align}
\langle S_{ {\bm e}_1 - {\bm e}_N } \rangle &=\sum_{1 \le i \neq j \le N} 
  e^{  { b_i -b_j}    }Z^{\text{5d}}_{1\text{-loop}} ({\bm p}={\bm e}_i-{\bm e}_j) 
+Z_{\text{mono}} ({\bm p}={\bm 0}, {\bm B}={\bm e}_1-{\bm e}_N ).  
\label{eq:adjexp}
\end{align}

The monopole bubbling effect in \eqref{eq:adjexp} is evaluated by the brane construction and localization formula for the elliptic genus. 
As explained in Section \ref{sec:section41}, the brane configuration for  the monopole bubbling effect ${\bm B}={\bm e}_1-{\bm e}_N$ and   ${\bm p}=0$ is depicted in Figure \ref{fig:screen1}. 
The quiver diagram of the bubbling GLSM is specified  by Figure \ref{fig:quiver1}. 
From the localization formula, the  JK residue for
the elliptic genus of the GLSM in Figure \ref{fig:quiver1}  is same as the following contour integral: 
\begin{align}
 Z^{(\eta)}_{\text{mono}}({\bm p}={\bm 0}, {\bm B}={\bm e}_1-{\bm e}_N )    = \text{sign}(\eta) \frac{2 \eta(\tau)^3 \vartheta_1(   \epsilon)}
{ \vartheta_1 (\pm m_{\text{ad}} +   \frac{1}{2}\epsilon) } \oint {d u} 
  \prod_{i=1}^{N}  \frac{\vartheta_1 (\pm ( u -a_i)-m_{\text{ad}} )}
{\vartheta_1( \pm ( u -a_i) +\frac{1}{2}  \epsilon) }.
\label{eq:mbubbling11}
\end{align}
Here $\text{sign}(\eta)$ is $+1$ for $\eta >0$ and $-1$ for $\eta <0$, respectively.
When $\eta >0$, the JK residue operation is  the  residues  at poles $ u = a_i -\frac{1}{2}  \epsilon$ for $i=1,\cdots,N$. 
On the other hand, when $\eta <0$, the JK residue operation is  the residues  at poles $ u = a_i +\frac{1}{2}  \epsilon$ for $i=1,\cdots,N$. The elliptic genus for the
the monopole bubbling effect is given by
\begin{align}
 Z^{(\eta>0)}_{\text{mono}} ({\bm p}={\bm 0}, {\bm B}={\bm e}_1-{\bm e}_N ) = 
\sum_{i=1}^N \prod_{j=1 \atop j \neq i}^{N}  \frac{\vartheta_1 (  a_i -a_j-m_{\text{ad}}-\frac{1}{2} \epsilon   ) \vartheta_1 (  a_j -a_i-m_{\text{ad}}+  \frac{1}{2} \epsilon  ) }
{\vartheta_1(  a_i -a_j ) \vartheta_1(  a_j -a_i+  \epsilon)},  \\
 Z^{(\eta<0)}_{\text{mono}} ({\bm p}={\bm 0}, {\bm B}={\bm e}_1-{\bm e}_N ) = 
\sum_{i=1}^N \prod_{j=1 \atop j \neq i}^{N}  \frac{\vartheta_1 (  a_j -a_i-m_{\text{ad}}+\frac{1}{2} \epsilon   ) \vartheta_1 (  a_i -a_j-m_{\text{ad}}-  \frac{1}{2} \epsilon  ) }
{\vartheta_1(  a_i -a_j+  \epsilon) \vartheta_1(  a_j -a_i ) }.
\end{align}
Note that $Z^{(\eta>0)}_{\text{mono}}=Z^{(\eta<0)}_{\text{mono}}$. This follows from the fact that the sum of the residues of a rational form on the torus is zero.

By comparing the expressions of  $ 
\langle S_{ {\bm e}_1 } \rangle * \langle S_{-{\bm e}_N} \rangle$, $\langle S_{-{\bm e}_N} \rangle *
\langle S_{ {\bm e}_1 } \rangle$   and $\langle S_{ {\bm e}_1 - {\bm e}_N } \rangle$, we find that the algebraic relation of  the 't Hooft surface operators:
\begin{align}
\langle {S}_{{\bm e}_1-{\bm e}_N} \rangle   =\langle {S}_{{\bm e}_1} \rangle *  
 \langle {S}_{-{\bm e}_N} \rangle =\langle {S}_{-{\bm e}_N} \rangle *
\langle {S}_{{\bm e}_1} \rangle 
\end{align}
Therefore we find that a surface operator with a higher charge is generated by surface operators with the minimal charges.
Applying  the Weyl-Wigner transform to $\langle {S}_{{\bm e}_1} \rangle *  
 \langle {S}_{-{\bm e}_N} \rangle =\langle {S}_{-{\bm e}_N} \rangle *
\langle {S}_{{\bm e}_1} \rangle$, we obtain the commutativity of two elliptic Ruijsenaars operators  $[\widehat{S}_{{\bm e}_1},  
 \widehat{S}_{-{\bm e}_N}]=0$. 

%%%%%%%%%%%%%%%%%%%%%%%%%%%%%%
\rem{
The SUSY localization formula for $\langle S_{{\bm e}_1-{\bm e}_N} \rangle$  is given by
\begin{align}
\langle S_{{\bm e}_1-{\bm e}_N} \rangle &=\sum_{1 \le i \neq j \le N} 
  e^{  { b_i -b_j}    }Z^{\text{5d}}_{1\text{-loop}} ({\bm p}={\bm e}_i-{\bm e}_j) 
+Z_{\text{mono}} ({\bm p}={\bm 0}, {\bm B}={\bm e}_1-{\bm e}_N ) \nonumber \\
&=\sum_{1 \le i \neq j \le N} 
  e^{  { b_i -b_j}    } \nonumber \\
&+
\sum_{i=1}^N \prod_{j=1 \atop j \neq i}^{N}  \frac{\vartheta_1 (  a_i -a_j-m_{\text{ad}}+\frac{1}{2} \epsilon   ) \vartheta_1 (  a_j -a_i-m_{\text{ad}}-  \frac{1}{2} \epsilon  ) }
{\vartheta_1(  a_i -a_j ) \vartheta_1(  a_j -a_i-  \epsilon)}.  
\end{align}
\begin{align}
Z^{\text{5d}}_{1\text{-loop}} ({\bm p}={\bm e}_i-{\bm e}_j) 
&=\left(
 \prod_{s=\pm1} \frac{\vartheta_1(s (a_i -a_j)-m_{\text{ad}}+\frac{1}{2} \epsilon )\vartheta_1(s (a_i -a_j)-m_{\text{ad}}-\frac{1}{2} \epsilon )  }{\vartheta_1(s (a_i -a_j))\vartheta_1(s (a_i -a_j)+\epsilon)  }
\right. \nonumber \\
&\left. \qquad  
\times  \prod_{k=i,j}\prod_{l \neq i,j} \frac{  \vartheta_1(s (a_k -a_l)-m_{\text{ad}} )}{\vartheta_1(s (a_k -a_l)+\frac{1}{2}\epsilon)  } 
\right)^{\frac{1}{2}}
\end{align}
}
%%%%%%%%%%%%%%%%%%%%%%%%%%%%%%

%%%%%%%%%%%%%%%%%%%%%%%%%%%%%%%%%%%%%%%%%%%%
\subsection{Surface operator $S_{(1,1, 0, \cdots, 0, -1,-1 )}$}

\begin{figure}[thb]
\centering
\subfigure[]{\label{fig:monopole21}
\includegraphics[width=5cm]{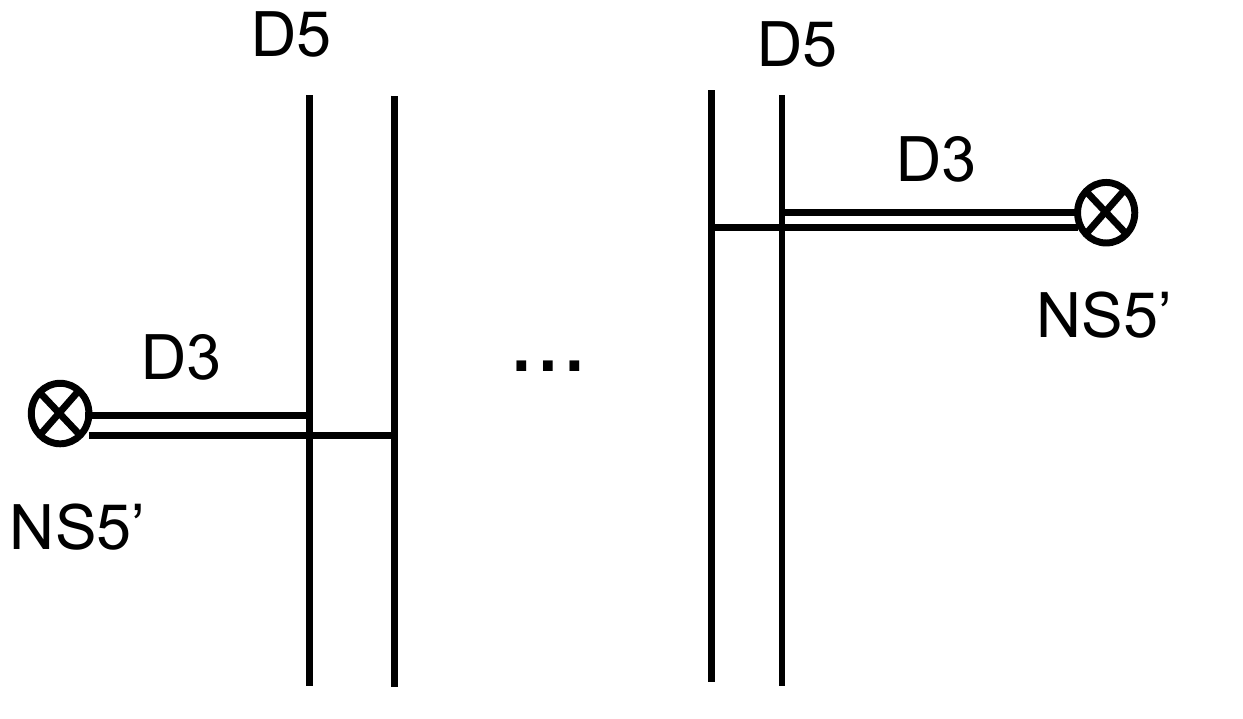}}
\subfigure[]{\label{fig:monopole22}
\includegraphics[width=5cm]{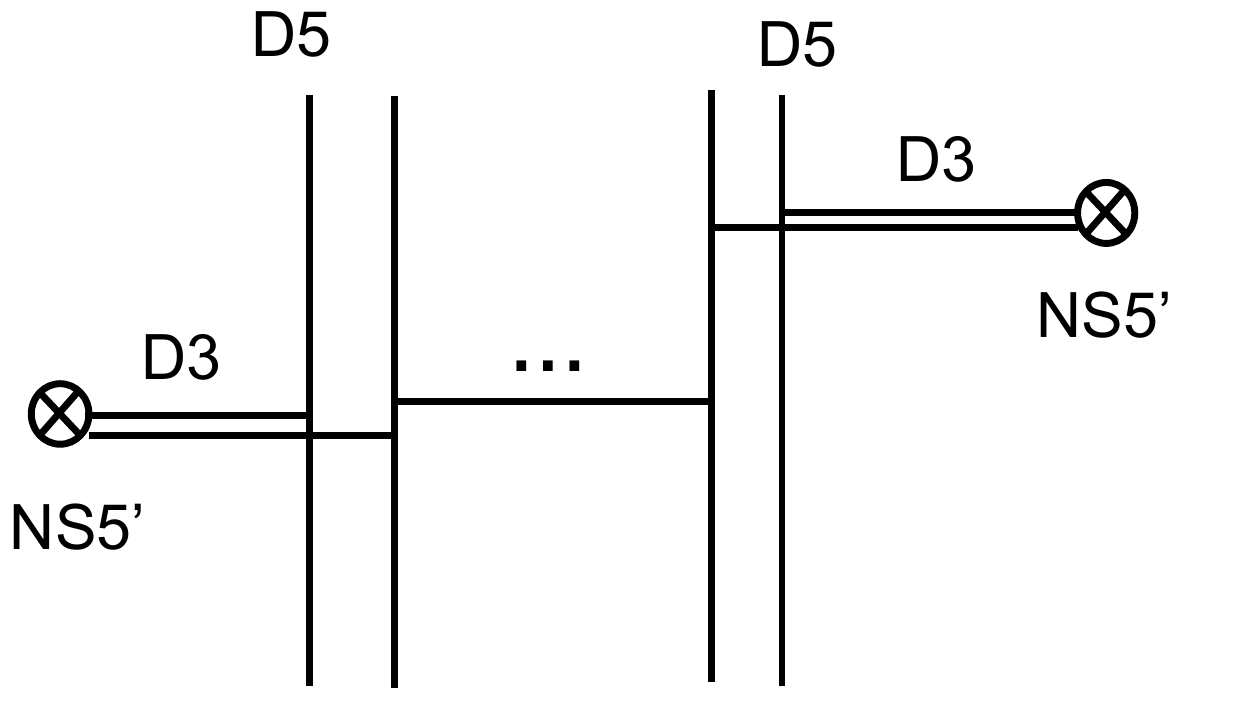}}
\subfigure[]{\label{fig:monopole23}
\includegraphics[width=3.5cm]{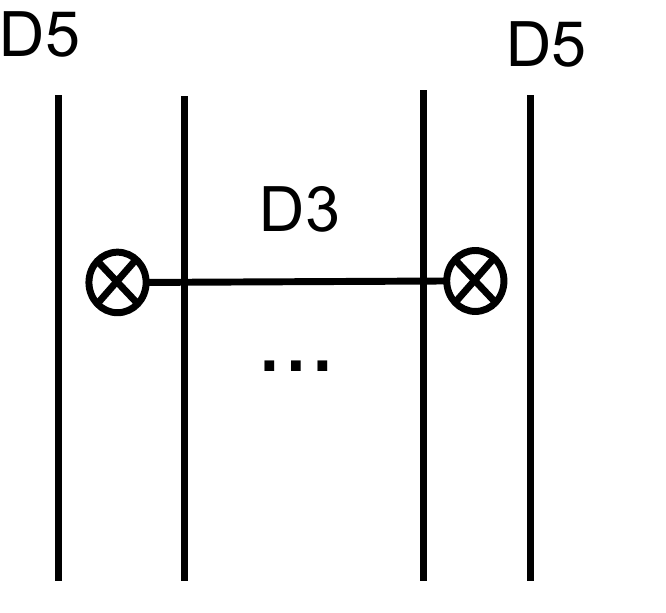}}
\caption{(a): The brane configuration for the surface operator with the magnetic charge $ {\bm B}={\bm e}_1+{\bm e}_2 -{\bm e}_{N-1}- {\bm e}_N$. 
 (b) A D3-brane suspended between the second D5-brane and the $(N-1)$-th D5-brane is introduced to describe the 't Hooft-Polyakov monopole with the charge ${\bm B}=-{\bm e}_2 + {\bm e}_{N-1}$. 
(c): After Hanany-Witten effect, we obtain the bran configuration for describing the  monopole bubbling effect ${\bm p}={\bm e}_1-{\bm e}_N$.}
\label{fig:screen3}
\end{figure}

Next we study the monopole bubbling effects in $\langle S_{(1,1, 0, \cdots, 0, -1,-1 )} \rangle =\langle S_{{\bm e}_1+{\bm e}_2-{\bm e}_{N-1}-{\bm e}_N} \rangle$, and study the algebra of 't Hooft surface operators.
The Moyal products of $\langle S_{{\bm e}_1+{\bm e}_2}  \rangle$ and $\langle S_{-{\bm e}_{N-1}-{\bm e}_N} \rangle$ are evaluated as
\begin{align}
&\langle S_{{\bm e}_1+{\bm e}_2}  \rangle *\langle S_{-{\bm e}_{N-1}-{\bm e}_N} \rangle
= 
 \sum_{1 \le i < j \le N, 1 \le k < l \le N  \atop \{ k, l \} \cap \{  i,j \}= \emptyset } 
  e^{   b_i +b_j-b_k- b_l    }Z^{\text{5d}}_{1\text{-loop}} ({\bm p}= {\bm e}_i+{\bm e}_j-{\bm e}_k-{\bm e}_l) \nonumber \\
&+\sum_{1 \le i \neq j \le N} 
  e^{  { b_i -b_j}    }Z^{\text{5d}}_{1\text{-loop}} ({\bm p}={\bm e}_i-{\bm e}_j) 
\sum_{l=1 \atop l \neq i,j}^N \prod_{k=1 \atop k \neq l}^{N}  \frac{\vartheta_1 (  a_l -a_k-m_{\text{ad}}+  \frac{1}{2} \epsilon  ) \vartheta_1 (  a_k -a_l-m_{\text{ad}}-\frac{1}{2} \epsilon   )  }
{\vartheta_1(  a_l -a_k+  \epsilon) \vartheta_1(  a_k -a_l ) }
 \nonumber \\
&\quad +
\sum_{i,j=1}^N \prod_{l=i,j} \prod_{k=1 \atop k \neq i,j}^{N}  \frac{\vartheta_1 (  a_l -a_k-m_{\text{ad}}+\frac{1}{2} \epsilon   ) \vartheta_1 (  a_k -a_l-m_{\text{ad}}-  \frac{1}{2} \epsilon  ) }
{\vartheta_1(  a_l -a_k + \epsilon) \vartheta_1(  a_k -a_l)},
\label{eq:secMoyal1}
\\
&\langle S_{-{\bm e}_{N-1}-{\bm e}_N} \rangle *
\langle S_{ {\bm e}_1+{\bm e}_2 } \rangle=
 \sum_{1 \le i < j \le N, 1 \le k < l \le N  \atop \{ k, l \} \cap \{  i,j \}= \emptyset } 
  e^{   b_i +b_j-b_k- b_l    }Z^{\text{5d}}_{1\text{-loop}} ({\bm p}= {\bm e}_i+{\bm e}_j-{\bm e}_k-{\bm e}_l) \nonumber \\
&+\sum_{1 \le i \neq j \le N} 
  e^{  { b_i -b_j}    }Z^{\text{5d}}_{1\text{-loop}} ({\bm p}={\bm e}_i-{\bm e}_j) 
\sum_{l=1 \atop l \neq i,j}^N \prod_{k=1 \atop k \neq l}^{N}  \frac{\vartheta_1 (  a_l -a_k-m_{\text{ad}}-\frac{1}{2} \epsilon   ) \vartheta_1 (  a_k -a_l-m_{\text{ad}}+  \frac{1}{2} \epsilon  ) }
{\vartheta_1(  a_l -a_k) \vartheta_1(  a_k -a_l+  \epsilon)}
 \nonumber \\
&\quad +\sum_{i,j=1}^N \prod_{l=i,j} \prod_{k=1 \atop k \neq i,j}^{N}  \frac{\vartheta_1 (  a_l -a_k-m_{\text{ad}}-\frac{1}{2} \epsilon   ) \vartheta_1 (  a_k -a_l-m_{\text{ad}}+ \frac{1}{2} \epsilon  ) }
{\vartheta_1(  a_l -a_k ) \vartheta_1(  a_k -a_l+  \epsilon)}.
\label{eq:secMoyal2}
\end{align}

We evaluate $\langle S_{{\bm e}_1+{\bm e}_2-{\bm e}_{N-1}-{\bm e}_N} \rangle$ and compare it with the Moyal products \eqref{eq:secMoyal1} and \eqref{eq:secMoyal2}.
The localization formula for $\langle S_{{\bm e}_1+{\bm e}_2-{\bm e}_{N-1}-{\bm e}_N} \rangle$ has the following expression.
\begin{align}
\langle S_{{\bm e}_1+{\bm e}_2-{\bm e}_{N-1}-{\bm e}_N} \rangle &= 
\sum_{1 \le i < j \le N, 1 \le k < l \le N  \atop \{ k, l \} \cap \{  i,j \}= \emptyset } 
  e^{   b_i +b_j-b_k- b_l    }Z^{\text{5d}}_{1\text{-loop}} ({\bm p}= {\bm e}_i+{\bm e}_j-{\bm e}_k-{\bm e}_l) \nonumber \\
&+\sum_{1 \le i \neq j \le N} 
  e^{  { b_i -b_j}    }Z^{\text{5d}}_{1\text{-loop}} ({\bm p}={\bm e}_i-{\bm e}_j) Z_{\text{mono}} ({\bm p}= {\bm e}_i- {\bm e}_j) \nonumber \\
&\quad + Z_{\text{mono}} ({\bm p}= {\bm 0} ) .
\label{eq:adjoint2}
\end{align}
Here we suppressed ${\bm B}={\bm e}_1+{\bm e}_2-{\bm e}_{N-1}-{\bm e}_N$ in $Z_{\text{mono}}$'s to shorten the expression.
The one-loop determinants $Z^{\text{5d}}_{1\text{-loop}} $ are  given by \eqref{eq:1loop1} and \eqref{eq:1loop3}. In this case, there are two monopole bubbling sectors specified by ${\bm p}={\bm e}_i-{\bm e}_j$ and ${\bm p}= {\bm 0}$.

%%%%%%%%%%%%%%%%%%%%%%%%%%%
\rem{
From the formula \eqref{eq:lopptot}, the one-loop determinant are given
\begin{align}
&Z^{\text{5d}}_{1\text{-loop}} ({\bm p}= {\bm e}_i+{\bm e}_j-{\bm e}_k-{\bm e}_l) \nonumber \\
&=
\left(
\prod_{h=i,j} \prod_{q=k,l}  \frac{\vartheta_1(\pm (a_h-a_q)-m_{\text{ad}}\pm \frac{\epsilon}{2})}{\vartheta_1(\pm (a_h-a_q)+\epsilon)\vartheta_1(\pm (a_h-a_q))  } 
\cdot \prod_{h=i,j, k,l} \prod_{q \neq i,j,k,l}  \frac{\vartheta_1(\pm (a_h-a_q)-m_{\text{ad}})}{\vartheta_1(\pm (a_h-a_q)+\frac{\epsilon}{2})  }
\right)^{\frac{1}{2}}
\end{align}
}
%%%%%%%%%%%%%%%%%%%%%%%%%%%
First let us evaluate the monopole bubbling effect specified by ${\bm p}={\bm e}_i-{\bm e}_j$. The brane configuration for the 't Hooft surface operator with the magnetic charge  ${\bm B}={\bm e}_1+{\bm e}_2-{\bm e}_{N-1}-{\bm e}_N$ is depicted by Figure \ref{fig:monopole21}. We introduce 
a D3-brane suspended between two D5-branes as Figure \ref{fig:monopole22}, which correspond to an 
't Hooft-Polyakov monopole. 
After Hanany-Witten transition, the matter contents of the low energy world volume theory on the D3-brane is  that the gauge group is $U(1)$ and the number of $\mathcal{N}=(4,4)$ hypermultiplets is $N-2$.  For example the brane configuration for $i=1, j=N$  depicted by Figure \ref{fig:monopole23}. Then  
 the monopole bubbling effect for ${\bm B}={\bm e}_1+{\bm e}_2-{\bm e}_{N-1}-{\bm e}_N$ and 
${\bm p}={\bm e}_i-{\bm e}_j$  is given by
\begin{align}
 Z^{(\eta)}_{\text{mono}}({\bm p}={\bm e}_i-{\bm e}_j )    = \text{sign}(\eta) \frac{2 \eta(\tau)^3 \vartheta_1(   \epsilon)}
{ \vartheta_1 (\pm m_{\text{ad}} +   \frac{1}{2}\epsilon) } \oint {d u} 
  \prod_{k=1 \atop k \neq i, j}^{N}  \frac{\vartheta_1 (\pm ( u -a_k)-m_{\text{ad}} )}
{\vartheta_1( \pm ( u -a_k) +\frac{1}{2}  \epsilon) }.
\label{eq:mbubbling111}
\end{align}

\eqref{eq:mbubbling111} is evaluate in similar way as \eqref{eq:mbubbling11}.
\begin{align}
 Z^{(\eta>0)}_{\text{mono}} ({\bm p}= {\bm e}_i- {\bm e}_j)   = 
\sum_{l=1 \atop l \neq i,j}^N \prod_{k=1 \atop k \neq l}^{N}  \frac{\vartheta_1 (  a_l -a_k-m_{\text{ad}}-\frac{1}{2} \epsilon   ) \vartheta_1 (  a_k -a_l-m_{\text{ad}}+  \frac{1}{2} \epsilon  ) }
{\vartheta_1(  a_l -a_k) \vartheta_1(  a_k -a_l+  \epsilon)},  \\
 Z^{(\eta<0)}_{\text{mono}} ({\bm p}= {\bm e}_i- {\bm e}_j)   = 
\sum_{l=1 \atop l \neq i,j}^N \prod_{k=1 \atop k \neq l}^{N}  \frac{\vartheta_1 (  a_l -a_k-m_{\text{ad}}+  \frac{1}{2} \epsilon  ) \vartheta_1 (  a_k -a_l-m_{\text{ad}}-\frac{1}{2} \epsilon   )  }
{\vartheta_1(  a_l -a_k+  \epsilon) \vartheta_1(  a_k -a_l ) }.
\end{align}
Again we have $ Z^{(\eta>0)}_{\text{mono}} ({\bm p}= {\bm e}_i- {\bm e}_j)= Z^{(\eta<0)}_{\text{mono}} ({\bm p}= {\bm e}_i- {\bm e}_j)$.

\begin{figure}[thb]
\centering
\subfigure[]{\label{fig:monopole31}
\includegraphics[width=5cm]{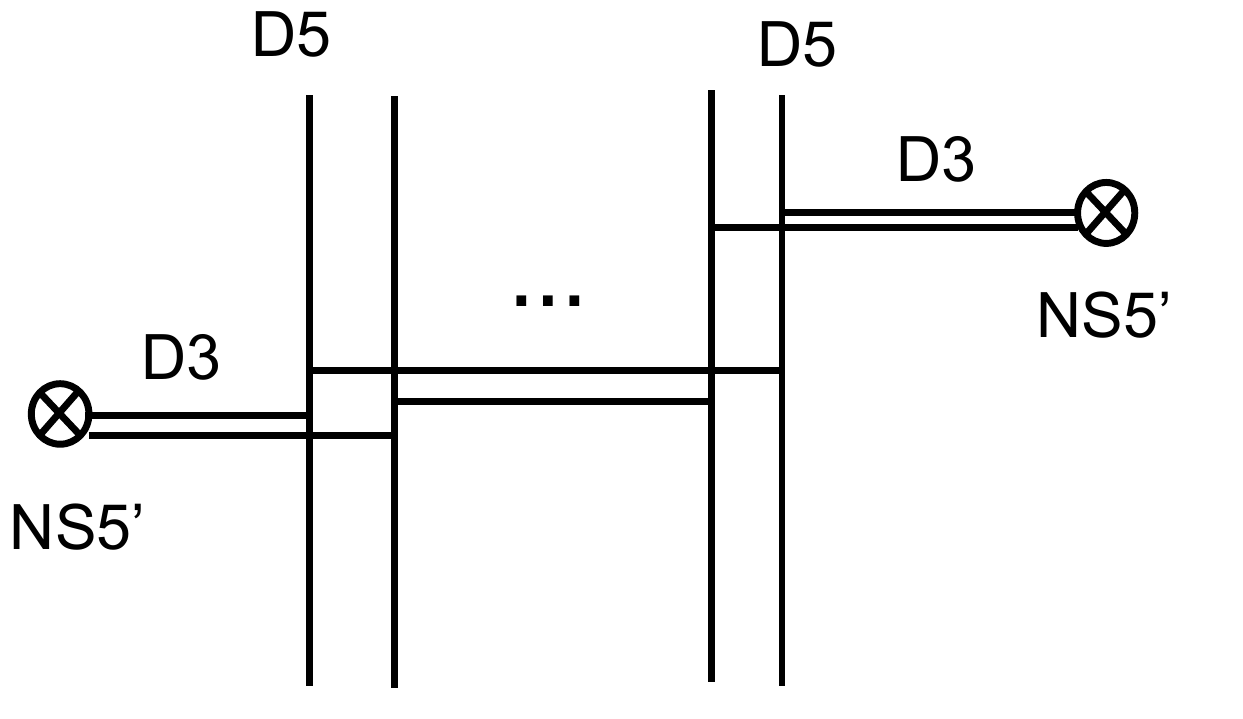}}
%\hspace{1cm}
\subfigure[]{\label{fig:monopole32}
\includegraphics[width=5cm]{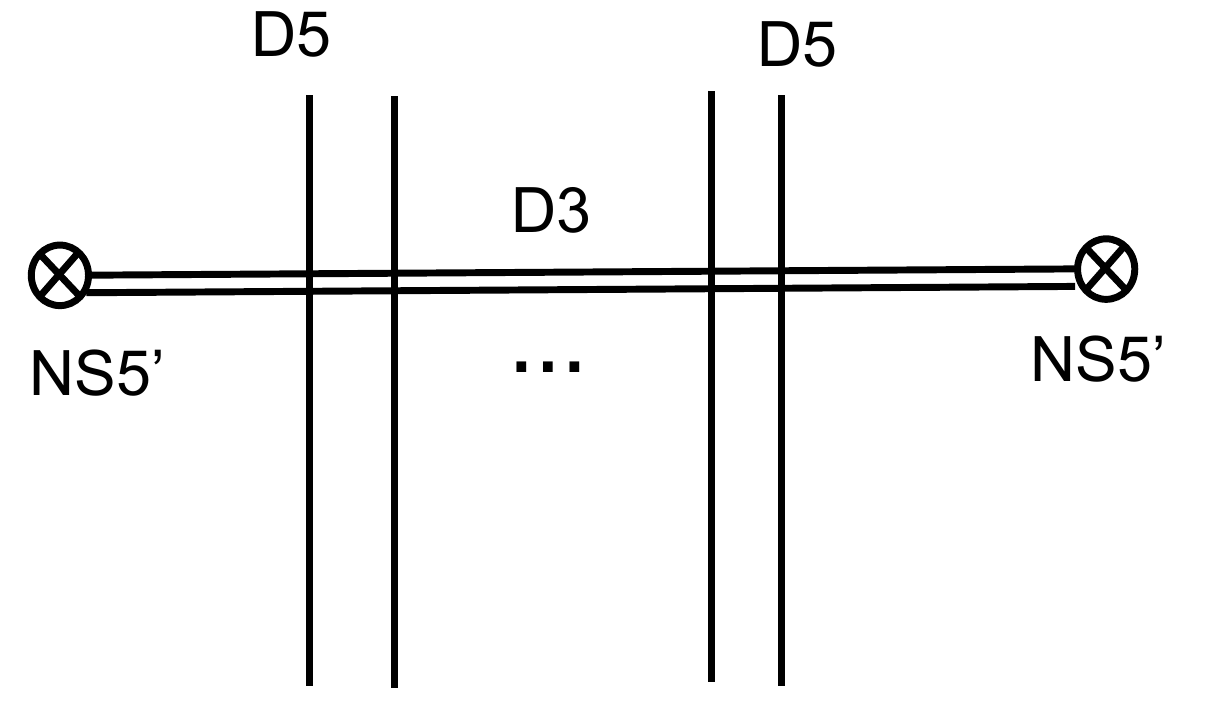}}
\subfigure[]{\label{fig:D3quiver2}
%\hspace{-5mm}
\includegraphics[width=1cm]{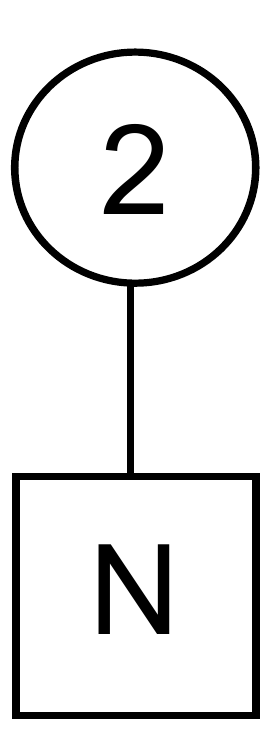}}
\caption{(a): We added two D3-branes suspended between D5-branes to Figure \ref{fig:monopole21}. 
 (b) When the positions of the segments of  D3-branes coincide, the screening of monopole charges charge  occurs. 
The world volume theory on two D3-branes give the monopole bubbling effect for ${\bm p}={\bm 0}$.  
(c): The quiver diagram denotes the matter content of the low energy world volume theory on the D3-branes of Figure \ref{fig:monopole32}. The circle with $2$ denotes  $\mathcal{N}=(4,4)$ $U(2)$ vector multiplet, and the solid line 
denotes the $N$ hypermultiplets in the fundamental representation of $U(2)$.}
\label{fig:screen4}
\end{figure}

Next we evaluate the monopole bubbling effect specified by ${\bm p}={\bm 0}$ from the brane construction. To achieve the monopole bubbling effect, we introduce 
two D3-branes suspended between D5-branes depicted as Figure \ref{fig:monopole31}. 
When the positions of the D3-branes coincides, we obtain the brane configuration in Figure \ref{fig:monopole32}. The low energy world volume theory on  two D3-branes in \ref{fig:monopole32} is   
 the 2d $\mathcal{N}=(4,4)$ $U(2)$ GLSM   with $N$ hypermultiplets. The localization formula for the elliptic genus is given by
\begin{align}
Z^{(\eta)}_{\text{mono}} ({\bm p}= {\bm 0} ) &=\frac{\left( 2 \pi \eta(\tau)^3 \right)^{2}}{2} 
\sum_{u_{\ast} \in \mathfrak{M}_{\text{sing}}  }\mathop{\text{JK-Res}}_{ u = {u}_*} ({Q}_*,{\eta})    \prod_{1 \le a \neq b \le 2} \vartheta_1  ( u_a-u_b ) 
   \nonumber \\
& \times 
 \prod_{ a,b=1 }^2  \frac{  \vartheta_1( u_a-u_b+\epsilon) }{  \vartheta_1 (u_a-u_b \pm {m}_{\text{ad}} + \frac{1}{2}\epsilon)}   \cdot \prod_{ b=1 }^2 \prod_{ i=1 }^N \frac{\vartheta_1 (\pm ( u_b -a_i)-m_{\text{ad}})}
{\vartheta_1( \pm ( u_b -a_i) +\frac{1}{2}  \epsilon) } d u_1 \wedge du_2\,.
\end{align}
When a vector $\eta$ is proportional to $(1,1)$, From the definition of the JK residues, 
$\mathfrak{M}_{\text{sing}}$ in the $(u_1,u_2)$ space, where JK residues are evaluated 
are specified by the intersection point of the following singular hyperplanes:
\begin{align}
&\left\{ u_1 -a_i+\frac{1}{2}  \epsilon =0 \right\} \cap \left\{ u_2 -a_j+\frac{1}{2}  \epsilon =0 \right\} \text{ for } i, j=1,\cdots, N, 
\label{eq:intersec1} \\
&\left\{ u_2-u_1 \pm {m}_{\text{ad}} +  \frac{1}{2}\epsilon \right\} \cap \left\{ u_1 -a_i+\frac{1}{2}  \epsilon =0 \right\} 
\text{ for } i=1,\cdots, N,
\label{eq:intersec2}
\\
&\left\{ u_1-u_2  \pm  {m}_{\text{ad}} +\frac{1}{2}\epsilon \right\} \cap \left\{ u_2 -a_i+\frac{1}{2}  \epsilon =0 \right\}  
\text{ for } i=1,\cdots, N\,.
\label{eq:intersec3}
\end{align}
The JK residues at the intersection points of \eqref{eq:intersec2} and \eqref{eq:intersec3}  are zero and the monopole bubbling effect is given by the JK 
residue at the intersection points of  \eqref{eq:intersec1} as
\begin{align}
 &Z^{(\eta=(1,1))}_{\text{mono}} ({\bm p} ={\bm 0}, {\bm B}={\bm e}_1-{\bm e}_N ) 
 \nonumber \\
& \qquad = 
\sum_{i,j=1}^N \prod_{l=i,j} \prod_{k=1 \atop k \neq i,j}^{N}  \frac{\vartheta_1 (  a_l -a_k-m_{\text{ad}}-\frac{1}{2} \epsilon   ) \vartheta_1 (  a_k -a_l-m_{\text{ad}}+ \frac{1}{2} \epsilon  ) }
{\vartheta_1(  a_l -a_k ) \vartheta_1(  a_k -a_l+  \epsilon)}.
\end{align}
In the similar way, when $\eta$ is proportional to (-1,-1),  
JK residues are evaluated at 
\begin{align}
&\left\{ -(u_1 -a_i)+\frac{1}{2}  \epsilon =0 \right\} \cap \left\{ -(u_2 -a_j)+\frac{1}{2}  \epsilon =0 \right\} \text{ for } i, j=1,\cdots, N, 
\label{eq:intersec11} \\
&\left\{ u_1-u_2  \pm {m}_{\text{ad}} + \frac{1}{2}\epsilon \right\} \cap \left\{-( u_1 -a_i)+\frac{1}{2}  \epsilon =0 \right\} 
\text{ for } i=1,\cdots, N,
\label{eq:intersec22}
\\
&\left\{ u_2-u_1\pm {m}_{\text{ad}}  + \frac{1}{2}\epsilon \right\} \cap \left\{ -(u_2 -a_i)+\frac{1}{2}  \epsilon =0 \right\}  
\text{ for } i=1,\cdots, N\,.
\label{eq:intersec33}
\end{align}
the monopole bubbling effect are given by 
\begin{align}
& Z^{(\eta=(-1,-1))}_{\text{mono}} ({\bm p}={\bm 0}, {\bm B}={\bm e}_1-{\bm e}_N )
\nonumber \\
& \qquad  = 
\sum_{i,j=1}^N \prod_{l=i,j} \prod_{k=1 \atop k \neq i,j}^{N}  \frac{\vartheta_1 (  a_l -a_k-m_{\text{ad}}+\frac{1}{2} \epsilon   ) \vartheta_1 (  a_k -a_l-m_{\text{ad}}-  \frac{1}{2} \epsilon  ) }
{\vartheta_1(  a_l -a_k + \epsilon) \vartheta_1(  a_k -a_l)}.
\end{align}
For the same reason as before,  $Z^{(\eta=(1,1))}_{\text{mono}}=Z^{(\eta=(-1,-1))}_{\text{mono}}$.
By comparing the Moyal products  with the SUSY localization formula for \eqref{eq:adjoint2}, we find the algebraic relation between vevs of 't Hooft surface operators:
\begin{align}
 \langle S_{{\bm e}_1+{\bm e}_2-{\bm e}_{N-1}-{\bm e}_N} \rangle=\langle S_{{\bm e}_1+{\bm e}_2}  \rangle *\langle S_{-{\bm e}_{N-1}-{\bm e}_N} \rangle=\langle S_{-{\bm e}_{N-1}-{\bm e}_N} \rangle *
\langle S_{ {\bm e}_1+{\bm e}_2 } \rangle.
\label{eq:secmoyal4}
\end{align}
By applying the Weyl-Wigner transform to \eqref{eq:secmoyal4}, we obtain the commutation relation of elliptic Ruijsenaars operators:
  $[\widehat{S}_{ {\bm e}_1+{\bm e}_2 }, \widehat{S}_{-{\bm e}_{N-1}-{\bm e}_N}]=0$.

%%%%%%%%%%%%%%%%%%%%%%%%%%%%%%%%%%%%
\section{Discussion}
\label{sec:section7}
We discuss the results in our paper and  the future directions. 
We  introduced magnetically charged surface operators on $T^2 \times \mathbb{R}^3$ and evaluated the expectation values in terms of supersymmetric localization. The SUSY localization formula obtained in this paper gives an elliptic deformation of localization formula for BPS 't Hooft loops on $S^1 \times \mathbb{R}^3$ \cite{Ito:2011ea} and the one for BPS bare monopole operators in \cite{Okuda:2019emk, Assel:2019yzd}. We  concentrated on  5d $\mathcal{N}=1^*$ $U(N)$ gauge theory and concretely studied the algebra of surface operators and monopole bubbling effects with small magnetic charges. 
  For general 5d $\mathcal{N}=1$  gauge theories, monopole bubbling effects are described  by elliptic genera of 2d $\mathcal{N}=(0,4)$ GLSMs,  which are obtained by the T-dual picture of the brane configuration studied  in \cite{Hayashi:2020ofu}. The gauge anomaly cancellation condition for 2d $\mathcal{N}=(0,4)$ GLSMs  give  constraints on the matter contents of the five dimensional theory and also magnetic charge ${\bm B}$. 
So far we do not have a clear understanding of the role of  gauge anomaly cancellation condition from the five-dimensional view point.

In Section \ref{sec:section5}, we found that the deformation quantization of 't Hooft surface operators in 5d $\mathcal{N}=1^*$ gauge theory agrees with the type-A elliptic  Ruijsenaars operators. 
Although the integrable structure appears in the 't Hooft surface operators is not manifest in the supersymmetric gauge theory, 
by using dualities in string theory, the  brane configuration for 't Hooft   operators without the monopole bubbling effect  is interpreted as  defects  in  
four-dimensional Chern-Simons theory, where the quantum integrable structure  naturally appear  \cite{Maruyoshi:2020cwy}.

In Section \ref{sec:section6}, we studied monopole bubbling effects and algebraic relation of 
't Hooft surface operators.  The physical interpretation of the Moyal product and Weyl-Wigner transform are as follows.
For 't Hooft loop operator on $S^1 \times \mathbb{R}^2_{\epsilon} \times \mathbb{R} $, the Moyal product of vevs of $n$ 't Hooft loops is identified with $n$-point correlation functions of 't Hooft loop operators \cite{Ito:2011ea} \cite{Hayashi:2019rpw}. Then we expect  a similar result holds, i.e., 
\begin{align}
\langle S_{{\bm B}_1} \cdot S_{{\bm B}_2}  \cdots S_{{\bm B}_{n-1}} \cdot S_{{\bm B}_n} \rangle= \langle S_{{\bm B}_1} \rangle * \langle S_{{\bm B}_2} \rangle * \cdots 
* \langle S_{{\bm B}_{n-1}} \rangle * \langle S_{{\bm B}_n} \rangle.
\label{eq:nmoya}
\end{align}
Here the center of Dirac monopole for the surface operator $S_{{\bm B}_i}$ for $i=1,\cdots, n$  locates at $(x^1,x^2,x^3)=(0,0,x^3_i)$ with $x^3_1 > x^3_2 > \cdots > x^3_n$. 
If an algebraic relation we find is written as $\langle S_{{\bm B_1}+{\bm B_2}}\rangle=\langle S_{{\bm B_1}}\rangle* \langle S_{{\bm B_2}}\rangle$, \eqref{eq:nmoya}  implies that
 the operator product expansion of 't Hooft surface operators $S_{{\bm B}_2} S_{{\bm B}_2}=S_{{\bm B_1}+{\bm B_2}} $ holds in  the correlation function.
It is desirable to evaluate correlation functions for surface operators in terms of supersymmetric localization and to establish the 
relation between the correlation functions and the Moyal products of vevs of a single 't Hooft surface operator.

Another future  direction is to study an elliptic deformation of the Coulomb branch chiral rings for three  dimensional gauge theories.
In 3d $\mathcal{N}=4$ gauge theories, the algebra of   the BPS bare monopole operators, the coulomb branch scalars and  the BPS dressed monopole operators is defined in terms of  the Moyal product and the Weyl-Wigner transform  \cite{Okuda:2019emk}. Then it found that the algebra agrees with the Coulomb branch chiral ring  and its deformation quantization (a.k.a quantized Coulomb branch) in \cite{Bullimore:2015lsa}.  
  Here the three dimensional version of algebraic relations studied in Section \ref{sec:section6} are identified with  ring relations of bare monopole operators in the quantized Coulomb branch.
 Thus the algebraic relation we find  is expected to be related to  ring relations in an elliptic deformation of the  quantized  Coulomb branches.
 It is interesting to study the five-dimensional uplifts of the Coulomb branch scalars and the dressed monopole operators  and define an elliptic deformation of the quantized Coulomb branches.

%%%%%%%%%%%%%%%%%%%%%%%%%%%%%%%%%%%%%%%%%
\section*{Acknowledgements}
The author would like to thank  Hirotaka~Hayashi, Kazunobu~Maruyoshi, Hiraku~Nakajima, and Takuya~Okuda for helpful discussions.

\appendix
\section{Gamma matrices}
\label{sec:appendix1}
\rem{
$32 \times 32$ gamma matrices satisfying the relation
\begin{align}
\{{\bm \Gamma}^{M}, {\bm \Gamma}^N \}=2 \delta^{M N}
\end{align}
${\bm \Gamma}^{M}$ is written as
\begin{align}
{\bm \Gamma}^{M}=
\left(
    \begin{array}{cc}
       0 & \tilde{\Gamma}^M \\
      \Gamma^M & 0 
    \end{array}
  \right)
\end{align}
}
%%%%%%%%%%%%%%%
 We summarize the definition of gamma matrices and their useful properties.
The $16 \times 16$ gamma matrices $\Gamma^M$ are defined as follows. 
\begin{align}
{ \Gamma}^{M}&=
\left(
    \begin{array}{cc}
       0 & E^T_{M+1} \\
      E_{M+1} & 0 
    \end{array}
  \right), \quad M = 1, 2, 3, 5, 6, 7, \\
{ \Gamma}^{4}&=
\left(
    \begin{array}{cc}
       0 & E^T_1 \\
      E_1 & 0 
    \end{array}
  \right), \quad
{ \Gamma}^{8}=
\left(
    \begin{array}{cc}
       0 & E^T_5 \\
      E_5 & 0 
    \end{array}
  \right),
 \\
{ \Gamma}^{9}&=
\left(
    \begin{array}{cc}
       1_{8 \times 8} & 0 \\
      0 & 1_{8 \times 8} 
    \end{array}
  \right), \quad
%\\
{ \Gamma}^{0}=
\left(
    \begin{array}{cc}
       {\rm i}1_{8 \times 8} & 0 \\
      0 & {\rm i} 1_{8 \times 8} 
    \end{array}
  \right).
\end{align}
Here $E_i$ for $i=1,\cdots, 8$ are defined by
\begin{align}
E_{\mu}&=
\left(
    \begin{array}{cc}
       J_{\mu} & 0 \\
       0& \bar{J}_{\mu}  
    \end{array}
  \right), \,\, \mu=1, 2, 3, 4, \quad
%  \\
E_{A}=
\left(
    \begin{array}{cc}
     0  & -J^T_A \\
      {J}_A & 0
    \end{array}
  \right), A=5,6,7,8.
\end{align}
with
{\scriptsize{
\begin{align}
(J_1,J_2,J_3,J_4)&=
\left( \left(
    \begin{array}{cccc}
       1 & 0 & 0 &0 \\
       0 & 1 & 0 &0\\  
       0 &0 & 1 &0 \\
       0 & 0& 0& 1
    \end{array}
  \right), \left(
    \begin{array}{cccc}
       0 & -1 & 0 &0 \\
       1 & 0 & 0 &0\\  
       0 &0 & 0 &-1 \\
       0 & 0& 1& 0
    \end{array}
  \right), \left(
    \begin{array}{cccc}
       0  & 0 & -1 &0 \\
        0 & 0 &0 & 1  \\  
       1&0 &0 & 0  \\
        0 & -1 & 0& 0
    \end{array}
  \right), \left(
    \begin{array}{cccc}
       0  & 0 &  0 & -1 \\
        0 & 0 &-1 & 0  \\  
       0&1 &0 & 0  \\
        1 & 0 & 0& 0
    \end{array}
  \right) \right),
\\
(\bar{J}_1,\bar{J}_2,\bar{J}_3,\bar{J}_4)&=
\left( \left(
    \begin{array}{cccc}
       1 & 0 & 0 &0 \\
       0 & 1 & 0 &0\\  
       0 &0 & 1 &0 \\
       0 & 0& 0& 1
    \end{array}
  \right), \left(
    \begin{array}{cccc}
       0 & -1 & 0 &0 \\
       1 & 0 & 0 &0\\  
       0 &0 & 0 &1 \\
       0 & 0& -1& 0
    \end{array}
  \right), \left(
    \begin{array}{cccc}
       0  & 0 & -1 &0 \\
        0 & 0 &0 & -1  \\  
       1&0 &0 & 0  \\
        0 & 1 & 0& 0
    \end{array}
  \right), \left(
    \begin{array}{cccc}
       0  & 0 &  0 & -1 \\
        0 & 0 &1 & 0  \\  
       0&-1 &0 & 0  \\
        1 & 0 & 0& 0
    \end{array}
  \right) \right), \\
(J_5,J_6,J_7,J_8)&=
\left( \left(
    \begin{array}{cccc}
       1 & 0 & 0 &0 \\
       0 & -1 & 0 &0\\  
       0 &0 & -1 &0 \\
       0 & 0& 0& -1
    \end{array}
  \right), \left(
    \begin{array}{cccc}
       0 & 1 & 0 &0 \\
       1 & 0 & 0 &0\\  
       0 &0 & 0 &1 \\
       0 & 0& -1& 0
    \end{array}
  \right), \left(
    \begin{array}{cccc}
       0  & 0 & 1 &0 \\
        0 & 0 &0 & -1  \\  
       1&0 &0 & 0  \\
        0 & 1 & 0& 0
    \end{array}
  \right), \left(
    \begin{array}{cccc}
       0  & 0 &  0 & 1 \\
        0 & 0 &1 & 0  \\  
       0&-1 &0 & 0  \\
        1 & 0 & 0& 0
    \end{array}
  \right) \right)\,.
\end{align}
}}

Another gamma matrix $\tilde{\Gamma}^M$ is defined by
\begin{align}
&\tilde{\Gamma}^M
=\left\{\
\begin{array}{ll}
\displaystyle
-\Gamma^0 & \text{ for } M=0 , \\
\displaystyle
\Gamma^M   & \text{ for } M=1,2,\cdots,9.
\end{array}
\right. 
\label{eq:tildeGamma}
\end{align}
Note that ${\Gamma}^M$ and $\tilde{\Gamma}^M$ satisfy the relation
\begin{align}
&   \tilde{ \Gamma}^M \Gamma^{N}+   \tilde{ \Gamma}^N \Gamma^{M}  =2 \delta^{M N}
, \quad \Gamma^{M}  \tilde{ \Gamma}^N+ \Gamma^{N}  \tilde{ \Gamma}^M  =2 \delta^{M N}, \\
&(\Gamma^{M})^T=\Gamma^{M}, \quad (\tilde{\Gamma}^{M})^T=\tilde{\Gamma}^{M}. 
\end{align}

$\Gamma^{M N}$, $\tilde{\Gamma}^{M N}$ and $\Gamma^{KLMN}$ are defined by
\begin{align}
\Gamma^{M N}&=\tilde{\Gamma}^{[M } \Gamma^{N]}, \quad \tilde{\Gamma}^{M N}=\Gamma^{[M } \tilde{\Gamma}^{N]}\,, \\
\Gamma^{KLMN}&=\tilde{\Gamma}^{[K } \Gamma^{L} \Gamma^{M} \Gamma^{N]}=\frac{1}{4 !} \sum_{\sigma \in \mathfrak{S}_4} \mathrm{sgn}(\sigma) \tilde{\Gamma}^{\sigma(K) } \Gamma^{\sigma(L)} \Gamma^{\sigma(M)} \Gamma^{\sigma(N)}
\end{align}
where $\mathfrak{S}_4$ is the permutation group of four elements and $\text{sgn} (\sigma)$ is is the signature of $\sigma \in \mathfrak{S}_4$. 
Here $[K L ]$ and $[K L M N ]$ denote the anti-symmetrization of   products of gamma matrices under the exchanges of any two indices.

%%%%%%%%%%%%%%%%%%%%%%%%%%%%%%%%%%%%%%%%%%
\section{One-loop determinants in 5d $\mathcal{N}=1^*$ gauge theory}
\label{App:1loop}
We summarize the explicit forms of the one-loop determinants of the surface operators studied in Section \ref{sec:section6}.  
From the localization computation of the one-loop determinant \eqref{eq:lopptot}, the one-loop determinants $Z^{\text{5d}}_{1\text{-loop}} ({\bm p})$ with 
${\bm p}={\bm e}_i-{\bm e}_j,$ and $ {\bm e}_i+{\bm e}_j-{\bm e}_k-{\bm e}_l$ in 5d $\mathcal{N}=1^*$ $U(N)$ gauge theory are given by 
\begin{align}
\label{eq:1loop1}
&Z^{\text{5d}}_{1\text{-loop}} ({\bm p}={\bm e}_i-{\bm e}_j) 
\nonumber \\
& \qquad=\left(
  \frac{\vartheta_1(\pm (a_i -a_j)-m_{\text{ad}}\pm \frac{1}{2} \epsilon )  }{\vartheta_1(\pm (a_i -a_j))\vartheta_1(\pm (a_i -a_j)+\epsilon)  }
 \prod_{k=i,j}\prod_{l \neq i,j}^N \frac{  \vartheta_1(\pm (a_k -a_l)-m_{\text{ad}} )}{\vartheta_1(\pm (a_k -a_l)+\frac{1}{2}\epsilon)  } 
\right)^{\frac{1}{2}} \,, \\
&Z^{\text{5d}}_{1\text{-loop}} ({\bm p}= {\bm e}_i+{\bm e}_j-{\bm e}_k-{\bm e}_l) \nonumber \\
&=
\left(
\prod_{h=i,j} \prod_{q=k,l}  \frac{\vartheta_1(\pm (a_h-a_q)-m_{\text{ad}}\pm \frac{\epsilon}{2})}{\vartheta_1(\pm (a_h-a_q)+\epsilon)\vartheta_1(\pm (a_h-a_q))  } 
\cdot \prod_{h=i,j, k,l} \prod_{q \neq i,j,k,l}  \frac{\vartheta_1(\pm (a_h-a_q)-m_{\text{ad}})}{\vartheta_1(\pm (a_h-a_q)+\frac{\epsilon}{2})  }
\right)^{\frac{1}{2}} \,.
\label{eq:1loop3}
\end{align}

%%%%%%%%%%%%%%%%%%%%%%%%%%%%%%%%%%%%%%%%%
\section{The differential operator $D_{1 0}$}
\label{ap:diff}
In order to compute the differential operator $D_{10}$, 
we explicitly write \eqref{eq:Qexacttot}.
\begin{align}
 V&= 
 (\Psi, \overline{ \widehat{\sf Q} \cdot \Psi}) +V_{\text{g.f}}  \nonumber \\
&=\int_{T^2 \times \mathbb{R}^3 } d^5 x \mathrm{Tr} \left[\sum_{M=1}^9 (\widehat{\sf Q} \cdot X_{0,M}-\bar{D}_{M} X_{1,8} ) \tilde{\Gamma}^{M} {\varepsilon}+{\rm i} \sum_{j=1}^7 X_{1,j} \nu^{j} \right] \nonumber \\
& \times  \overline{\left[  \bar{D}_{M} X_{0, N} (  \bar{\nu}_j {\Gamma}^{ M  N} \varepsilon  )\nu_{j}  +{\rm i}K_j \nu_{j} +   (2\bar{D}_M X_{0,4}-\bar{D}_M  \widehat{\sf Q} \cdot X_{1,8}-\bar{D}_4 X_{0,M}-{\rm i}\bar{D}_0 X_{0,M})  \tilde{\Gamma}^M {\varepsilon} \right]} 
\nonumber \\
&
+ X_{1,9} (\sum_{i=1,2,3,9} \bar{D}_i X_{0,i} +\frac{\xi}{2} \widehat{\sf Q} \cdot X_{1,9}) \nonumber \\
&=\int_{T^2 \times \mathbb{R}^3 } d^5 x \mathrm{Tr} \left( \left[\sum_{M=1}^9 (\widehat{\sf Q} \cdot X_{0,M}-\bar{D}_{M} X_{1,8} ) \tilde{\Gamma}^{M} {\varepsilon}+{\rm i} \sum_{j=1}^7 X_{1,j} \nu^{j} \right]
\right. \nonumber \\
&  \qquad \qquad \times \left[ 2 \bar{D}_{M} X_{0, N} (\nu_i \widetilde{\Gamma}^{ M N} {\varepsilon} ) \nu^{j}- {\rm i} \widehat{\sf Q} \cdot X_{1,j} \nu^{j}  - 2\bar{D}_{{z}} X_{0, M} \tilde{\Gamma}^M {\varepsilon} \right] \nonumber \\
&\left. \qquad \qquad + X_{1,9} ({\rm i} \sum_{i=1,2,3,9} \bar{D}_i X_{0,i} +\frac{\xi}{2} \widehat{\sf Q} \cdot X_{1,9}) \right) \nonumber \\
&= \int_{T^2 \times \mathbb{R}^3} \mathrm{Tr} \left[ \sum_{M=1}^9 ( \widehat{\sf Q} \cdot X_{0,M}- \bar{D}_{M} X_{1,8} ) ( 2\bar{D}_M X_{0,4}-\bar{D}_M  \widehat{\sf Q} \cdot X_{1,8}-\bar{D}_4 X_{0,M}-{\rm i}\bar{D}_0 X_{0,M})  \right. \nonumber \\
&\left. +  \sum_{j=1}^7 X_{1,j} ( \widehat{\sf Q} \cdot X_{1,j}+2 {\rm i} ({\nu}_j \widetilde{\Gamma}^{ M N} {\varepsilon}) \bar{D}_{M} X_{0, N}  ) 
+ X_{1,9} \left(\sum_{i=1,2,3,9} \bar{D}_i X_{0,i} +\frac{\xi}{2} \widehat{\sf Q} \cdot X_{1,9}
\right) \right]\,.
\label{eq:Vexplicit}
\end{align}
From \eqref{eq:Vexplicit}, the action of $D_{10}$ on  the  fluctuation fields ${\bm X}_{0}$ are read off as 
\begin{align}
(D_{10} \cdot {\bm X}_{0})_i& =-2{\rm i} \sum_{j,k=1}^3 ({\nu}_i \widetilde{\Gamma}^{j k } {\varepsilon}) \bar{D}_{j} X_{0, k}+ 2{\rm i} \overline{D}_{i} X_{0,9} -2{\rm i} \bar{D}_9 X_{0,i} 
\quad  i \in \{1,2,3 \}
\,. \\
(D_{10} \cdot {\bm X}_{0})_i&=2 {\rm i} \sum_{j=1}^3 \sum_{k=5}^8 (\nu_i \tilde{\Gamma}^{j k} \varepsilon)
 \bar{D}_k X_{0, k}+ 2{\rm i} \sum_{k=5}^9 (\nu_i \tilde{\Gamma}^{9 k} \varepsilon) \bar{D}_9 X_{0,k} \quad 
i \in \{4, 5, 6, 7 \}\,, \\
(D_{10} \cdot {\bm X}_{0})_8& =\bar{D}_M 
( 2\bar{D}_M X_{0,4}-\bar{D}_4 X_{0,M}-{\rm i}\bar{D}_0 X_{0,M})\,, \\
(D_{10} \cdot {\bm X}_{0})_9& ={\rm i} \sum_{i=1,2,3,9} \bar{D}_i X_{0,i}\,.
 \end{align}

\label{appendix1}

%%%%%%%%%%%%%%%%%%%%%%%%%%%%%%%%%%%%%%%%

\bibliography{refs}

\end{document}